%% file: ma.tex
\documentclass{article}

\usepackage{arxiv}

\usepackage[utf8]{inputenc} 
\usepackage[T1]{fontenc}    
\usepackage[hidelinks]{hyperref}       
\usepackage{url}            
\usepackage{booktabs}       
\usepackage{amsfonts}       
\usepackage{nicefrac}       
\usepackage{microtype}      
\usepackage{graphicx}
\usepackage{doi}
\usepackage[labelfont=bf]{caption}

\usepackage{tikz}
\usepackage{circuitikz}
\usetikzlibrary{positioning,arrows,chains,matrix,scopes,fit,calc,decorations,decorations.pathreplacing,fpu}
\usepackage{siunitx}
\usepackage{mhchem}
\usepackage{tabulary}
\usepackage{multicol}
\usepackage{amssymb}
\usepackage{marvosym}
\usepackage{csquotes}
\usepackage{csvsimple}
\usepackage{placeins}

\input{./figures/definition_scalebar.tikz}

\newcommand{\subfigHorizontalSep}{0.25}
\newcommand{\subfigVerticalSep}{0.25}

\newcommand{\THz}{\si{\tera\hertz}}
\newcommand{\THzTDS}{\THz-TDS~}
\newcommand{\numAperture}{\text{NA}}
\newcommand{\fieldOfView}{\text{FoV}}

\DeclareSIUnit{\fps}{fps}

\definecolor{AlertOrange}{HTML}{ff7f37}
\definecolor{BlockBlue}{HTML}{003380}

\title{Real-time High Resolution THz Imaging 
	with a Fiber Coupled Photo Conductive Antenna and an Uncooled Microbolometer Camera}

\author{ 
	\href{https://orcid.org/0000-0003-0935-0253}{\includegraphics[scale=0.06]{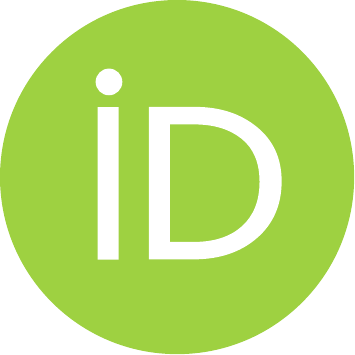}\hspace{1mm}Peter Zolliker\thanks{Correspondence; \texttt{peter.zolliker@empa.ch}}},   \href{https://orcid.org/0000-0002-6892-634X}{\includegraphics[scale=0.06]{orcid.pdf}\hspace{1mm}Elena Mavrona},  \href{https://orcid.org/0000-0003-1547-9889}{\includegraphics[scale=0.06]{orcid.pdf}\hspace{1mm}Erwin Hack}\\
	Transport at Nanoscale Interfaces Laboratory\\
	Empa, Swiss Federal Laboratories for Materials Science and Technology\\
	Überlandstrasse 129, 8600 Dübendorf, Switzerland\\	\\
	\And
	Mostafa Shalaby and Elisa Söllinger\\
	Swiss Terahertz Research-Zurich, Swiss Terahertz GmbH\\
	Technopark, 8005 Zurich, Switzerland and Park Innovaare, 5234 Villigen, Switzerland\\	
}

\renewcommand{\shorttitle}{THz Imaging with PCA and Microbolometer}

\hypersetup{
pdftitle={Real-time High Resolution THz Imaging 
	with a Fiber Coupled Photo Conductive Antenna and an Uncooled Microbolometer Camera},
pdfauthor={Peter Zolliker, Mostafa Shalaby, Elisa Söllinger, Elena Mavrona, Erwin Hack},
pdfkeywords={THz imaging;
	real-time;
	photo conductive antenna; 
	microbolometer camera; 
	THz-TDS},
}

\begin{document}
\maketitle

\begin{abstract}
	We present a real-time THz imaging method using a commercial fiber coupled photo conductive antenna as the THz source and an uncooled microbolometer camera for detection. Compared to other THz imaging setups, this concept is very adaptable due to its compact and uncooled radiation source, whose fiber coupling allows for a flexible placement. Using a camera with high sensitivity renders real-time imaging possible.  
	As a proof-of-concept, the beam shape of a THz Time Domain Spectrometer was measured.
	We also demonstrate the potential for practical applications in transmission geometry, covering material science and security tasks. 
	The results suggest that hidden items, complex structures and moisture contents of (biological) materials can be resolved. We discuss the limits of the current setup, possible improvements, potential (industrial) applications and outline the feasibility of imaging in reflection geometry or extending it to multi-spectral imaging using band pass filters. 
\end{abstract}


\section{Introduction}
\input{./sections/pca-imaging_introduction_v1.tex}

\section{Setups and Methods}
\input{./sections/pca-imaging_setups-and-methods_v1.tex}

\newpage
\section{Results}
\input{./sections/pca-imaging_results_v1.tex}

\newpage
\section{Discussion}
\input{./sections/pca-imaging_discussion_v1.tex}

\section{Conclusions}
\input{./sections/pca-imaging_conclusion_v1.tex}

\subsection*{Supplementary Material}
	The following are available as supplementary material: Video S1: Real-time \THzTDS beam profiling, Video S2: Real-time rotation of metallic Siemens star, Video S3: Real-time THz imaging of a key, Video S4: Demonstration of laboratory equipment with concealed key, Video S5: Real-time moisture content resolution in leaves, Video S6: Demonstration of laboratory equipment with wet leaves, Video S7: Real-time rotation of thin wood sample.

\subsection*{Author contributions}
	
	Conceptualization, M. S. and P. Z.; RIGI imaging scheme prototype development: E. S.; Software, P. Z.; Validation, P. Z.; Formal Analysis, P. Z.; Investigation, M. S., P. Z.; Resources, M. S. and E. M.; Data Curation, P. Z.; Writing – Original Draft Preparation, E. H., P. Z., E. M. and M. S.; Writing – Review \& Editing, E. H., P. Z., E. M. and M. S.; Visualization, P. Z.; Supervision, P. Z.; Project Administration, P. Z.

\subsection*{Funding}
	P. Z. gratefully acknowledges funding from the Swiss National Science Foundation, SNF project 200021\_179061 / 1.

\subsection*{Acknowledgments}
	
	The authors would like to thank the following colleagues from Laboratory for Transport at Nanoscale Interfaces for their help: Jil Graf and Rolf Brönnimann, for the 3D printing and laser ablation, respectively, of some of the samples; Daniel Sacré:  LaTeX support (typesetting, figure inclusion, layout), setup sketches and editing of videos in supplementary materials.\\
	Additional thanks to Anselm Deninger, Toptica Photonics AG, Gräfelfing, Germany for providing insight into the property differences for PCAs of different generations.

\subsection*{Conflicts of interest}
	M. S. and E. S. declare to be affiliated with Swiss Terahertz GmbH, Park Innovaare, 5234 Villigen, Switzerland. Swiss Terahertz is the supplier of the RIGI camera and lens that were used in this work.

\bibliographystyle{abbrv}



\input{ma.bbl}
\end{document}

%% file: figures/definition_scalebar.tikz
\newcommand{\scalebarHeight}{1.5}
\newcommand{\scalebarDivisionStep}{5}
\newcommand{\scalebarDistanceToImage}{3.75}
\newcommand{\scalebarDistanceCenterLabelToBar}{2.5}

\newcommand{\horizontalScalebarTEMPLATE}[7]{
	\coordinate (scalebarStart) at ($(#1.#2 east)+(0,#3 mm)$); 
	\coordinate(scalebarEnd) at ($(#1.#2 west)+(0,#3 mm)$); 
	\path ($(scalebarStart)!0.5!(scalebarEnd)$)++(0,#4 mm) coordinate(scalebarLabelCenter); 
	
	\csvreader[ head to column names,%
	late after head=\xdef\aold{\mmX}\xdef\bold{\mmY}\xdef\Iminold{\Imin}\xdef\Imaxold{\Imax},%
	after line=\xdef\aold{\mmX}\xdef\bold{\mmY}\xdef\Iminold{\Imin}\xdef\Imaxold{\Imax}]%
	{#5}{}{%
		
		\pgfmathsetmacro\scalebarNumberOfSteps{\mmX/\scalebarDivisionStep} 
		\pgfmathsetmacro\scalebarNumberOfStepsBlack{\scalebarNumberOfSteps-#6}
		\pgfmathsetmacro\scalebarNumberOfStepsWhite{\scalebarNumberOfSteps-#7}
		\pgfmathsetmacro\scalebarPercentSteps{1/\scalebarNumberOfSteps} 
		
		\foreach \x in {0,2,...,\scalebarNumberOfStepsBlack}{
			\pgfmathsetmacro\percentHelperStart{\x*\scalebarPercentSteps}
			\pgfmathsetmacro\percentHelper{(\x+1)*\scalebarPercentSteps}
			\coordinate(rectangleStart) at ($(scalebarStart)!\percentHelperStart!(scalebarEnd)$);
			\coordinate (scalebarRectangleHelper) at ($(scalebarStart)!\percentHelper!(scalebarEnd)$);
			\coordinate(rectangleEnd) at ($(scalebarRectangleHelper)+(0,-\scalebarHeight mm)$);
			\fill[black,draw=black](rectangleStart)rectangle(rectangleEnd);
			\coordinate(blackLabel\x) at ($(rectangleStart)!0.5!(scalebarRectangleHelper)$);
		}
		
		\foreach \x in {1,3,...,\scalebarNumberOfStepsWhite}{
			\pgfmathsetmacro\percentHelperStart{\x*\scalebarPercentSteps}
			\pgfmathsetmacro\percentHelper{(\x+1)*\scalebarPercentSteps}
			\coordinate (scalebarRectangleHelper) at ($(scalebarStart)!\percentHelper!(scalebarEnd)$);
			\fill[white,draw=gray]($(scalebarStart)!\percentHelperStart!(scalebarEnd)$)rectangle($(scalebarRectangleHelper)+(0,-\scalebarHeight mm)$);
		}
		
		\node(scalebarLabel) at (scalebarLabelCenter) {\SI[round-mode=places,round-precision=1]{\mmX}{\milli\meter}};
		\draw[thick,->|,>=stealth] let \p1=(scalebarLabel.east),\p2=(scalebarStart) in (\x1,\y1)--(\x2,\y1);
		\draw[thick,->|,>=stealth] let \p1=(scalebarLabel.west),\p2=(scalebarEnd) in (\x1,\y1)--(\x2,\y1);
	}
} 

\newcommand{\verticalScalebarTEMPLATE}[7]{
	\coordinate (scalebarStart) at ($(#1.south #2)+(#3 mm,0)$); 
	\coordinate(scalebarEnd) at ($(#1.north #2)+(#3 mm,0)$); 
	\path ($(scalebarStart)!0.5!(scalebarEnd)$)++(#4 mm,0) coordinate(scalebarLabelCenter); 
	
	\csvreader[ head to column names,%
	late after head=\xdef\aold{\mmX}\xdef\bold{\mmY}\xdef\Iminold{\Imin}\xdef\Imaxold{\Imax},%
	after line=\xdef\aold{\mmX}\xdef\bold{\mmY}\xdef\Iminold{\Imin}\xdef\Imaxold{\Imax}]%
	{#5}{}{%
		
		\pgfmathsetmacro\scalebarNumberOfSteps{\mmY/\scalebarDivisionStep} 
		\pgfmathsetmacro\scalebarNumberOfStepsBlack{\scalebarNumberOfSteps-#6}
		\pgfmathsetmacro\scalebarNumberOfStepsWhite{\scalebarNumberOfSteps-#7}
		\pgfmathsetmacro\scalebarPercentSteps{1/\scalebarNumberOfSteps} 
		
		\foreach \y in {0,2,...,\scalebarNumberOfStepsBlack}{
			\pgfmathsetmacro\percentHelperStart{\y*\scalebarPercentSteps}
			\pgfmathsetmacro\percentHelper{(\y+1)*\scalebarPercentSteps}
			\coordinate(rectangleStart) at ($(scalebarStart)!\percentHelperStart!(scalebarEnd)$);
			\coordinate (scalebarRectangleHelper) at ($(scalebarStart)!\percentHelper!(scalebarEnd)$);
			\coordinate(rectangleEnd) at ($(scalebarRectangleHelper)+(-\scalebarHeight mm,0)$);
			\fill[black,draw=black](rectangleStart)rectangle(rectangleEnd);
		}
		\foreach \y in {1,3,...,\scalebarNumberOfStepsWhite}{
			\pgfmathsetmacro\percentHelperStart{\y*\scalebarPercentSteps}
			\pgfmathsetmacro\percentHelper{(\y+1)*\scalebarPercentSteps}
			\coordinate (scalebarRectangleHelper) at ($(scalebarStart)!\percentHelper!(scalebarEnd)$);
			\fill[white,draw=gray]($(scalebarStart)!\percentHelperStart!(scalebarEnd)$)rectangle($(scalebarRectangleHelper)+(-\scalebarHeight mm,0)$);
		}
		
		\node[rotate=-90](scalebarLabel) at (scalebarLabelCenter) {\SI[round-mode=places,round-precision=1]{\mmY}{\milli\meter}};
		\draw[thick,->|,>=stealth] let \p1=(scalebarLabel.east),\p2=(scalebarStart) in (\x1,\y1)--(\x1,\y2);
		\draw[thick,->|,>=stealth] let \p1=(scalebarLabel.west),\p2=(scalebarEnd) in (\x1,\y1)--(\x1,\y2);
	}
} 

\newcommand{\horizontalScalebarTop}[4]{
	\horizontalScalebarTEMPLATE{#1}{north}{\scalebarDistanceToImage}{\scalebarDistanceCenterLabelToBar}{#2}{#3}{#4}
} 

\pgfmathsetmacro\bottomScalebarDistanceToImage{\scalebarDistanceToImage-\scalebarHeight}
\pgfmathsetmacro\bottomScalebarDistanceCenterLabelToBar{\scalebarDistanceCenterLabelToBar+\scalebarHeight}
\newcommand{\horizontalScalebarBottom}[4]{
	\horizontalScalebarTEMPLATE{#1}{south}{-\bottomScalebarDistanceToImage}{-\bottomScalebarDistanceCenterLabelToBar}{#2}{#3}{#4}
} 

\newcommand{\verticalScalebarRight}[4]{
	\verticalScalebarTEMPLATE{#1}{east}{\scalebarDistanceToImage}{\scalebarDistanceCenterLabelToBar}{#2}{#3}{#4}
} 

%% file: sections/pca-imaging_introduction_v1.tex
		In material science, as well as in industrial and security applications, non-destructive testing of samples 
		is an important prerequisite. Non-ionizing \THz~radiation can be an option, since it can deliver sub-millimeter resolution. 
		Additionally, many materials have high transmissivity in this frequency range.
		A broad range of materials like plastics \cite{wietzke_applications_2007,wietzke_terahertz_2010,kuter_thz_2018,naftaly_industrial_2019},
		ceramics \cite{niijima_evaluation_2018,mikerov_analysis_2020},
		illicit drugs \cite{miles_terahertz_2007,davies_terahertz_2008,liu_detection_2010}, 
		\mbox{explosives \cite{davies_terahertz_2008,liu_detection_2010,meilhan_active_2011},} 
		wood \cite{todoruk_origin_2012,tanaka_applicability_2014,zolliker_extracting_2017-1},
		\mbox{paper \cite{banerjee_diagnosing_2008,federici_review_2012}}, 
		leaves \cite{meilhan_active_2011,gente_monitoring_2015}
		and blood \cite{jeong_characterization_2013,kulya_fast_2020}
		have been successfully studied with \THz~radiation. Also, a large number of security applications based on (sub-)THz \mbox{radiation \cite{davies_terahertz_2008}} have been proposed and some are commercially available \cite{davies_terahertz_2008,tzydynzhapov_new_2020}. 
		Despite the immense potential, application of \THz~outside research is currently far from being common. In theory, a THz transmission imaging setup can be made out of a single line source, a collimating lens and a pixel array camera. This simplistic setup is a promising candidate for industrial and security applications. However, the achievable resolution and image quality, respectively, are limited by the irradiation wavelength, the numerical aperture \numAperture~of all optical components, as well as by the camera properties (pixel size, sensitivity, etc.). To circumvent especially the limitations of the optical components, lens-less \mbox{imaging \cite{hack_comparison_2016,valzania_thz_2019-1}} can be an option.\\
		To date, the most commonly used sources  in the frequency range between \SI{0.2}{\tera\hertz} and \SI{4}{\tera\hertz} are far infrared (FIR) gas lasers, quantum cascade lasers (QCLs) and photo-conductive antennas (PCAs).
		
		The FIR gas lasers \cite{crocker_stimulated_1964} are based on a high power, mid-infrared \mbox{\ce{CO2}-laser} pumping a THz cavity.
		Their THz emission can be continuous-wave (cw), with the output power exceeding $\SI{150}{\milli\watt}$ at \mbox{\SI{2.52}{\tera\hertz} \cite{valzania_thz_2019-1}}. The output wavelength depends on the gas in the THz resonator.
		However, cw lasers are only emitting a single line and the stable operation can be challenging. Recently, the relatively compact THz QCLs started performing without cryostat, operating with a thermoelectric cooler \cite{bosco_thermoelectrically_2019} and at temperatures up to \SI{250}{\kelvin} \cite{khalatpour_high-power_2021}. In frequency combs operation, the bandwidth has been higher than an octave \cite{rosch_octave-spanning_2015}, but still it is limited to $\SI{1}{\tera\hertz}-\SI{6}{\tera\hertz}$ \cite{fujita_recent_2018}.
		Recently, the reported peak output power reached \SI{2}{\watt} (\SI{58}{\kelvin}, \SI{3.3}{\tera\hertz}, single mode) \cite{jin_phase-locked_2020}. Despite the promising progress, more research is required to achieve room temperature operation, larger bandwidth and higher power. 
		\newline PCAs combine many advantages of the above mentioned sources: They are compact, well established broadband sources with bandwidth up to \SI{6}{\tera\hertz} and \SI{90}{\decibel} dynamic range \cite{dietz_influence_2014}.
		Their performance is limited by the near-infrared (NIR) pump pulse, carrier lifetime and the chosen detector. 
		The majority of commercially available THz Time Domain Spectrometer (THz-TDS) use a PCA combined with off-axis parabolic mirrors (OAPMs) as a basis. Application of the compact and robust \THzTDS quickly spread from the first reported use case of water vapor absorption characterization \cite{exter_terahertz_1989} into other research disciplines, even including (art) conservation \cite{krugener_thz_2017,krugener_-site_2019,krugener_terahertz_2020} and archaeology \cite{fukunaga_terahertz_2010,jackson_terahertz_2015,krugener_terahertz_2020,mikerov_analysis_2020}. 
		So far, for \THzTDS imaging only prototypes of multipixel detectors \cite{li_63-pixel_2020} are reported; image acquisition requires a sequential scanning of the sample which cannot provide data in real-time. Nevertheless, scanning \THzTDS paved the way for adaption of THz imaging in industrial applications, e.\,g. lacquer thickness determination \cite{helmut_fischer_gmbh_non-destructive_2018,naftaly_industrial_2019}. Due to PCAs being widely available, THz imaging with them is very attractive. For example, Stantchev et al. used a PCA for real-time single pixel imaging \cite{stantchev_real-time_2020}. Their method to modulate the THz beam via a digital micromirror device retains the time domain capabilities of the THz-TDS, whilst still achieving a resolution of $32\times32$ pixels at \mbox{6 frames-per-second (fps)}. Per contra, their approach requires elaborate equipment, whereas we propose a method based on a simple transmission setup, using a PCA as source and exploiting the recent improvements of microbolometer cameras. Our approach can deliver much higher resolution and is more suitable for in-field (industrial) applications, but sacrifices the spectral information. 
		\newline In this paper, we give a short overview of the method, the camera properties, the setups and describe the data processing. 
		We recorded a THz beam shape in real-time and determined the spatial resolution with a Siemens star. 
		The suitability of the method for practical applications was demonstrated by imaging a key concealed in a paper envelope, the qualitative resolution of different water contents in leaves and imaging of annual rings in wood. 
		Finally, we discuss the limitations and possible improvements of the setup as well as suggest practical applications and future extensions.

%% file: sections/pca-imaging_setups-and-methods_v1.tex
\subsection{Camera and lens properties}
	For the experiments, a Swiss Terahertz RIGI camera and THz lens were used. Their specifications are found in table \ref{tab:cameraProperties} and \ref{tab:lensProperties}, respectively. \\[-0.0cm]
	\begin{table}[h!]
		\centering
		\input{./tables/2020_pca-imaging_table-camera-properties.tex}
		\vspace{0.0625cm}
		\caption{
			Technical specification of the camera. 
		}
		\label{tab:cameraProperties}
	\end{table}

	\vspace{-0.0cm}
	\begin{table}[h!]
		\centering
		\input{./tables/pca-imaging_setup-and-methods_table-lens-properties.tex}
		\vspace{0.0625cm}
		\caption{
			Technical specification of the lens. 
		}
		\label{tab:lensProperties}
	\end{table}
	\vspace{-0.0cm}
	The used camera RIGI S2x is a new prototype that is optimized for low-frequency imaging. This is achieved through an optimized detector structure to enhance the absorption of low-frequency THz radiation.

\subsection{Setups}
\begin{figure}[h!]
		\centering
			\includegraphics[width=\linewidth,trim=1.5cm 1cm 0.25cm 1.675cm, clip]{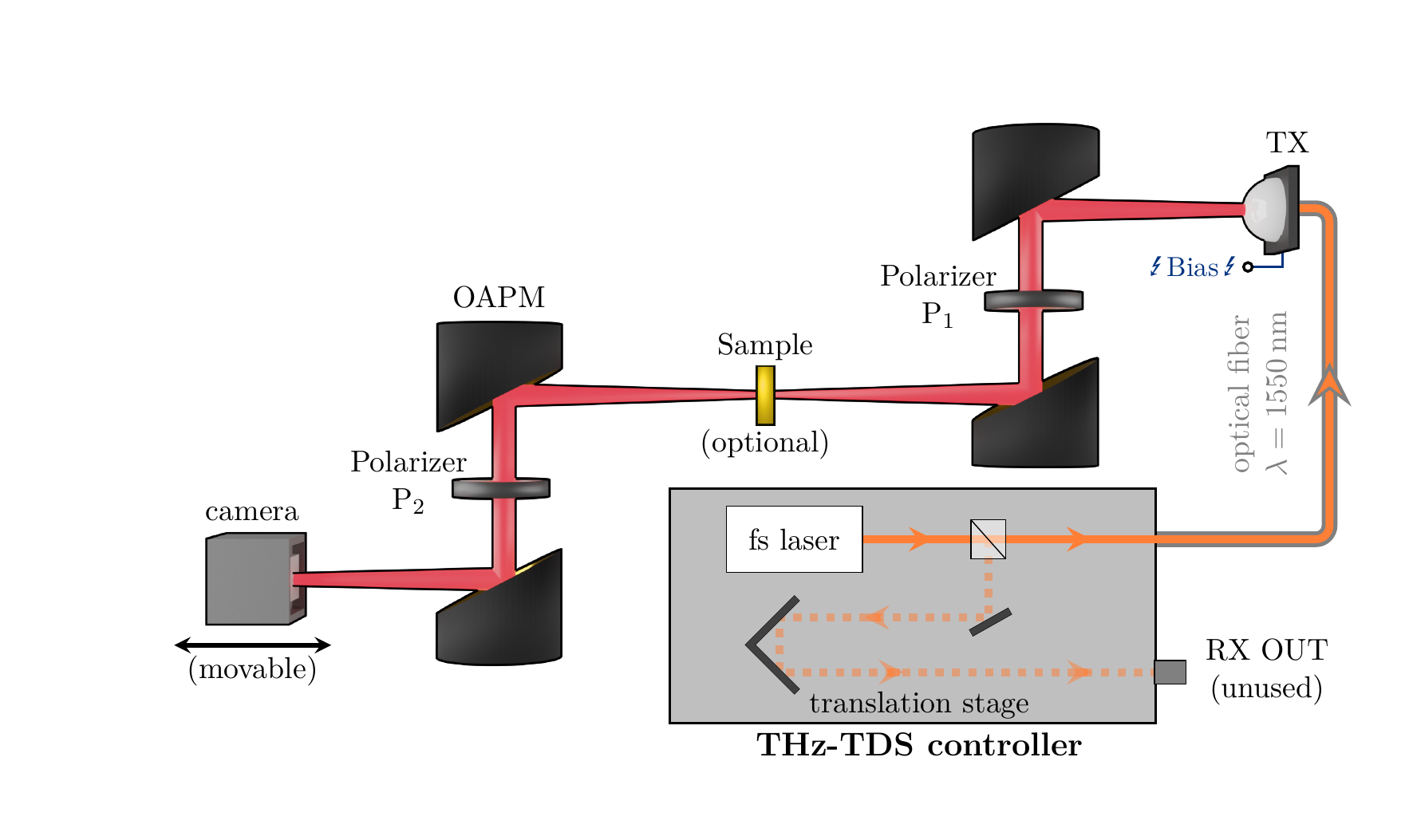}
		\caption{
			Schematic of the zigzag setup. Via an optical fiber, a fs pump laser ($\lambda=\SI{1550}{\nano\metre}$) excites the TX, which in turn emits THz radiation. Four OAPMs and two polarizers $\mathrm{P_1}$, $\mathrm{P_2}$ guide the THz emission onto the camera sensor (positioned where in a THz-TDS the RX would be located). 
		}
		\label{fig:TDSbeamshapeSetupSchematic}
	\end{figure}

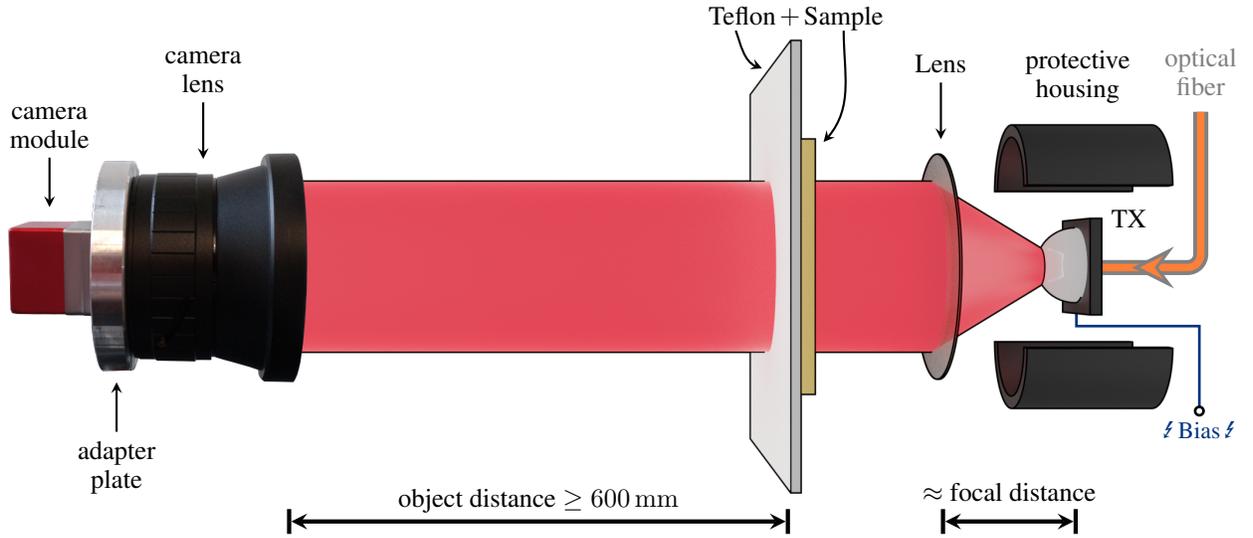
\begin{figure}[h!]
		\centering
		\input{./figures/setups/linearImaging/pca-imaging_setup-and-methods_fig2_linear-setup.tikz}
		\vspace{-0.375cm}
		\caption{
			Schematic of the lens-based imaging setup. The THz emission of the TX is collimated by a silicon lens before it reaches the sample. To suppress the thermal image, the sample is mounted onto a sheet of Teflon. The transmitted radiation is recorded with the camera/lens combination placed more than \SI{600}{\milli\metre} away from the object plane. Drawing not to scale.
		}
		\label{fig:linearImagingSetupSchematic}
	\end{figure}

	A commercial THz-TDS system (Tera-FlashTF-1503 Ver. 4 Dec/2015, Toptica Photonics AG, Gräfelfing, Germany) was used as a starting point. In this system, a \SI{100}{\micro\meter} InGaAs based strip-line antenna serves as transmitter (TX). It is biased with \SI{+120}{\volt} and optically pumped by a pulsed \SI{1550}{\nano\meter} Erbium fiber laser (pulse duration: \SI{60}{\femto\second}, repetition rate: \SI{100}{\mega\hertz}). The \SI{22.3}{\milli\watt} of NIR pump reaching the TX are converted to roughly \SI{40}{\micro\watt} cw equivalent, linearly polarized THz radiation. The THz-TDS scan time was kept fixed at \SI{70}{\pico\second} throughout all the experiments.\\ 
	The optical setup is of the zigzag-transmission geometry type (see \mbox{fig. \ref{fig:TDSbeamshapeSetupSchematic}}): An OAPM collimates the diverging output of the TX, which is then focused by another OAPM. A sample can be placed in the beam waist. The transmitted radiation is guided by a second OAPM pair (rotationally symmetrical to the first one) onto the detector. In a standard THz-TDS, the detector would be a receiver (RX), working on the inverted principle of the TX. In this work, the RX was replaced with an uncooled microbolometer camera, which could be displaced along the THz propagation direction. Since the sensor is sensitive to all emitted wavelengths, the spectral resolution is not recoverable from the data. On the other hand, a high real-time spatial resolution is achieved, in contrast to the single-pixel nature of the RX. 
	Furthermore, two wire-grid polarizers were inserted in the parallel beam sections
	to ensure a high degree of linear polarization. Additionally, they also allow for intensity reduction via the rotating polarizer method. To simplify the setup of \mbox{figure \ref{fig:TDSbeamshapeSetupSchematic}}, we  removed all the OAPMs, illuminated the sample directly and captured the image with a lens specifically designed for the RIGI camera (\mbox{fig. \ref{fig:linearImagingSetupSchematic}}).

\newpage
	For practical imaging applications, the complicated OAPM alignment is less favorable than the simpler lens-based setup depicted in \mbox{figure \ref{fig:linearImagingSetupSchematic}}. The TX is placed roughly in focus of a silicon (Si) lens ($f=\SI{25}{\milli\meter}$, $d=\SI{25}{\milli\meter}$), which collimates the divergent radiation of the THz emitter. The exact distance between lens and PCA determines the size of the illuminated area. The majority of the samples were mounted 
	close to the collimating lens on a \SI{1}{\milli\meter} thick sheet of Teflon for thermal image suppression. If this was not possible, a \SI{3}{\milli\meter} Teflon sheet was placed between sample and camera. Furthermore, black polyethylene (PE) foil was fixed to the TX 
	to weaken the leaking \SI{1550}{\nano\meter} NIR pump pulses. Due to the design of the camera lens ($f=\SI{44}{\milli\meter}$, $\text{f number}=\num{0.7}$), the minimum object distance is \SI{600}{\milli\meter}. 

\subsection{Image analysis and (post-)processing routines}
	\label{subsec:AnalysisAndPostRoutines}
	The pre-processed image data from the camera was sent via USB to a PC. A control software allowed for real-time filtering and the saving of the data in different compressed and loss-less file formats. For this publication, the data were saved as 14 bit integer with minimal filtering into loss-less csv-files; only once compressed 8 bit jpg was used. In MATLAB, the following post-processing (also suitable for a real-time data stream) was applied: At first, dead pixels were removed by replacing them with a neighboring pixel. Optionally, images were then filtered with a $3\times3$ median filter followed by a Gaussian filter with a width of $\sigma = 1$ pixel. 
	Then, a background image, captured while the THz beam was blocked and pre-processed with the same procedure, was subtracted.
	Afterwards, contrast enhancement was performed by re-scaling the minima/maxima of the gray-scale image data. For better visualization, some of the post-processed images were converted to false color.
	\newline Imaging of samples bigger than the illuminated area was made possible by scanning over the sample in real-time and stitching the single frames together. The position offset between the neighboring frames was determined by using auto-correlation in an area around the center of the image.
	\newline The spatial scaling of the THz images was estimated from the sensor pixel pitch and known feature dimensions measured on the samples. Sample (feature) dimensions were extracted with ImageJ/Fiji (see \mbox{e.\,g.} \cite{schneider_nih_2012,schindelin_fiji_2012}) from photographs of the samples on graph paper background.

%% file: tables/2020_pca-imaging_table-camera-properties.tex
\small
\begin{tabulary}{\linewidth}{lc}
		\toprule
			Camera & Swiss Terahertz RIGI S2x\\
		\hline
			Type &  uncooled THz microbolometer\\
			Operation range  & $\num{16}-\SI{3000}{\micro\metre} \,\,(0.1-\SI{18}{\tera\hertz})$\\
			Pixel size (\si{\micro\metre})  & 25\\
			Number of pixels  & $160\times120$\\
			Detector size (L$\,\times\,$H, \si{\milli\metre\squared}) & $4\times 3$ \\
			NEP 			
			& $<\SI{1.5}{\pico\watt}$ @ \SI{4.6}{\tera\hertz}	\\
			ADC (bit)  & 14 \\
			Frame transfer rate (fps) & 9\\
			Data transfer + power & USB \\
		\bottomrule
\end{tabulary}

%% file: tables/pca-imaging_setup-and-methods_table-lens-properties.tex
\small
\begin{tabulary}{\linewidth}{lc}
	\toprule
		Lens &  \\
	\hline
		Focal length (\si{\milli\metre}) & 44\\
		f number & 0.7\\
		Lens material & HRFZ-Si\\ 
		$\fieldOfView$ (lateral$\,\times\,$vertical) & $\SI{17.3}{\degree}\times\SI{13}{\degree}$\\
		Operation range & $\num{7.4}-\SI{3000}{\micro\metre}\,\,(\num{0.1}-\SI{30}{\tera\hertz})$\\
	\bottomrule
\end{tabulary}

%% file: figures/setups/linearImaging/pca-imaging_setup-and-methods_fig2_linear-setup.tikz
\begin{tikzpicture}[
		fsLaserInFibre/.style={line width = 0.5mm, double, double distance = 1mm,gray, double=AlertOrange,rounded corners=5pt}
]

		\newcommand{\propagationDirectionArrowFibreTEMPLATE}[9]{
			\draw[gray,line width=2 mm,#7,>={stealth[]},draw opacity=0] let \p1=(#1),\p2=(#2) in (\x1,\y2)++(#3)--++(#8);
			\draw[AlertOrange,line width=1 mm,#7,>={stealth[]},draw opacity=0] let \p1=(#1),\p2=(#2) in (\x1,\y2)++(#4)--++(#9);
		}

		\newcommand{\propagationDirectionArrowFibreHorizontalLeft}[2]{
			\propagationDirectionArrowFibreTEMPLATE{#1}{#2}{-2 mm, 0 mm}{-1 mm, 0 mm}{-2 mm, 0 mm}{-4 mm,0 mm}{->}{-1 mm,0 mm}{-1mm,0}
		}

		\node[anchor=south west,inner sep=0,outer sep=0] (linearImagingSetupBack) at (0,0){
			\includegraphics[width=0.9\textwidth,trim=0cm 0.75cm 0cm 1.125cm, clip]{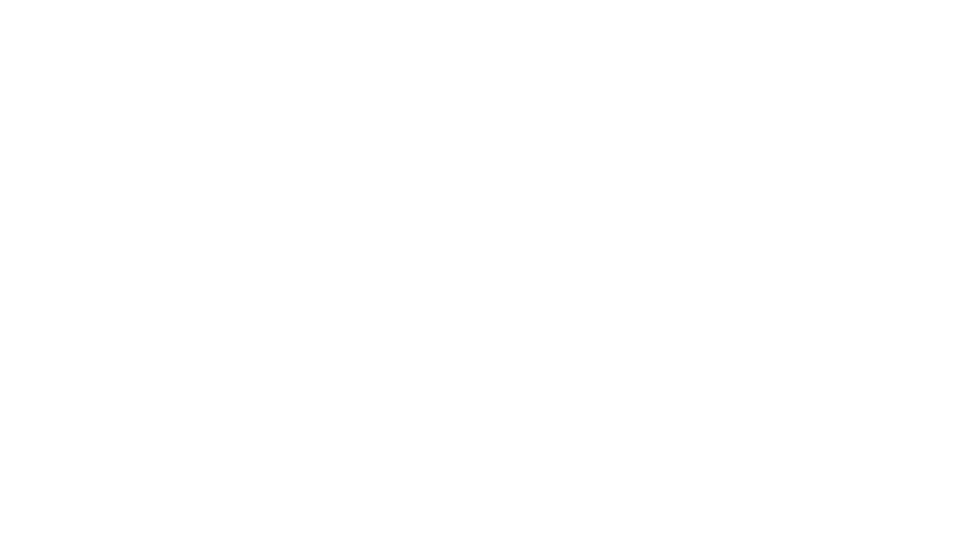}
		};

		\begin{scope}[x={(linearImagingSetupBack.south east)},y={(linearImagingSetupBack.north west)}]
					\coordinate (PCA) at (0.9075,0.51);
					\coordinate (housingLabelPos) at ($(PCA)+(0,0.32785)$);
					\coordinate(fiberEnd) at ($(PCA)+(0.11,-0.875mm)$);
					\coordinate(pumpLaserArrowHelper) at ($(PCA)!0.625!(fiberEnd)$);
					\coordinate (pumpLaserArrow) at ($(pumpLaserArrowHelper)+(0,-0.875mm)$);
					\coordinate(siLensSetupPos) at (0.785,0.725);

					\draw[fsLaserInFibre]($(PCA)+(0,-0.875mm)$)--($(fiberEnd)$)--++(0,0.275)node[above,text=gray,align=center](oFibreLabel){optical\\\phantom{gh}fiber\phantom{gh}};
					\propagationDirectionArrowFibreHorizontalLeft{pumpLaserArrow}{fiberEnd}
					\draw[BlockBlue,line width=1pt]($(PCA)+(-0.001255,-0.045)$)|-++(0.11,-0.07) to [short,-o]++(0,-0.15) node[below]{\small \Lightning \,Bias\,\Lightning}; 

							\path let \p1=(housingLabelPos), \p2=(oFibreLabel) in (\x1,\y2-0.25mm) coordinate(housingLabelPos2);
							\node[align=center,anchor=center] at (housingLabelPos2){protective\\housing};
							\node at ($(PCA)+(0.045755,0.075)$){TX};

							\node[align=center](sampleLabel) at (0.65,0.9375) {\phantom{p}$+$\phantom{p}};
							\node[anchor=east,inner sep=0,outer sep=0](teflonLabel) at ($(sampleLabel.west)+(0.25cm,0)$){\phantom{p}Teflon};
							\node[anchor=west,inner sep=0,outer sep=0](sampleLabel2) at ($(sampleLabel.east)+(-0.25cm,0)$){Sample};
							\draw[->,>=stealth,thick](sampleLabel2.south)to[in=45,out=-90](0.68,0.72);

							\draw[<-,>=stealth,thick](siLensSetupPos)--++(0,0.10)node[above,align=center]{Lens};

						\coordinate(objectDistanceCameraPos) at (0.2,0.05);
						\coordinate (objectDistanceObjectPos) at (0.65,0.05);
						\draw[ultra thick, |<->|,>=stealth](objectDistanceCameraPos)--(objectDistanceObjectPos);
						\node[above=0mm] at ($(objectDistanceCameraPos)!0.5!(objectDistanceObjectPos)$){object distance $\geq\SI{600}{\milli\meter}$};

						\draw[ultra thick, |<->|,>=stealth] let \p1=(siLensSetupPos), \p2=(PCA) in (\x1,0.05)coordinate(lensDistanceSetupLensPos)--(\x2,0.05)coordinate(lensDistancePCALensPos);
						\node[above=1.5mm, align=center] at ($(lensDistanceSetupLensPos)!0.5!(lensDistancePCALensPos)$){$\approx$ focal distance};

			\node[inner sep=0,outer sep=0] (linearImagingSetupBack) at (0.5,0.5){
				\includegraphics[width=0.9\textwidth,trim=0cm 0.75cm 0cm 0.75cm, clip]{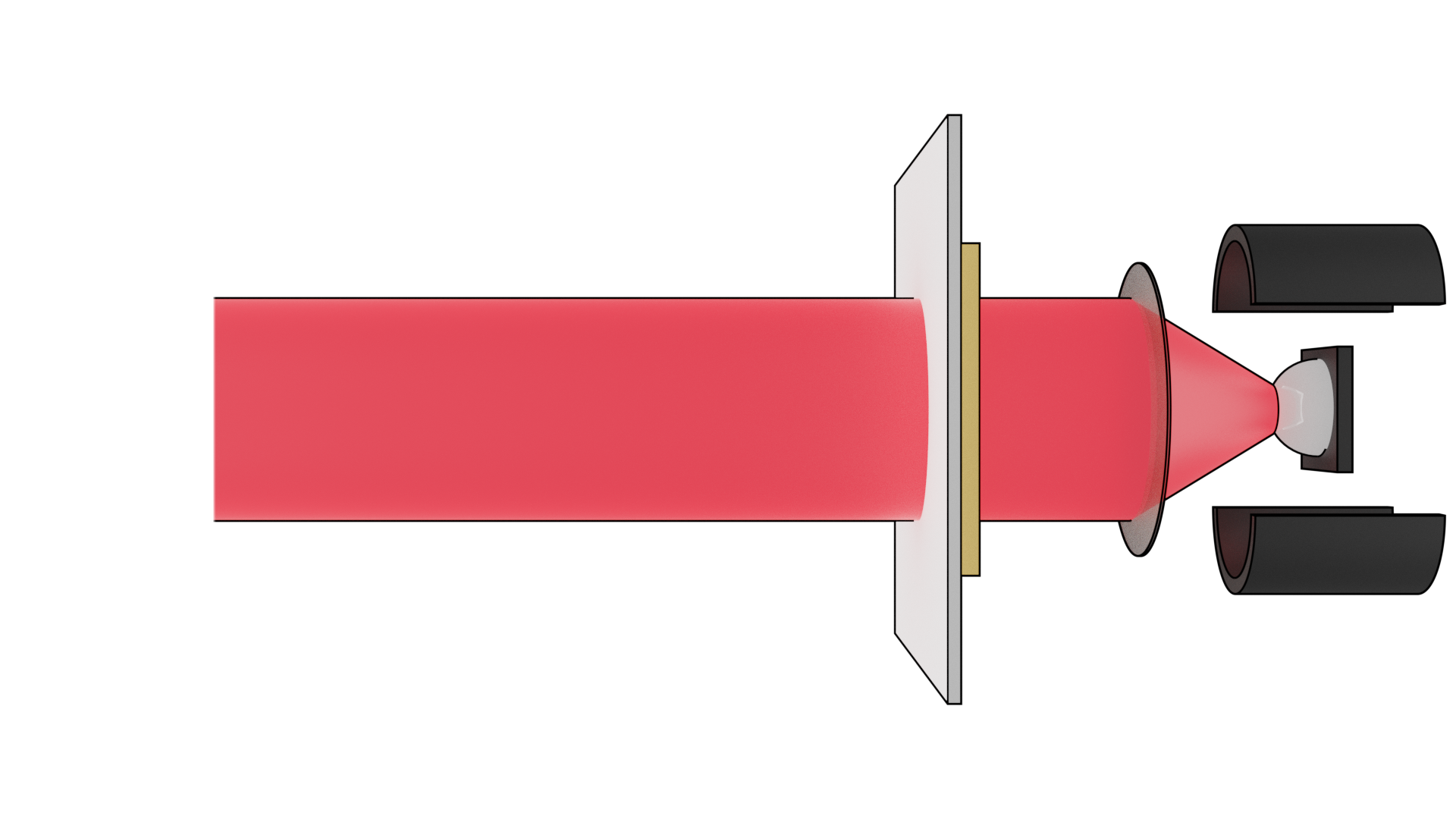}
			};

			\node[anchor=east,inner sep=0, outer sep=0] (linearImagingCameraCoverup) at (0.22,0.4975){
				\includegraphics[width=0.245\textwidth]{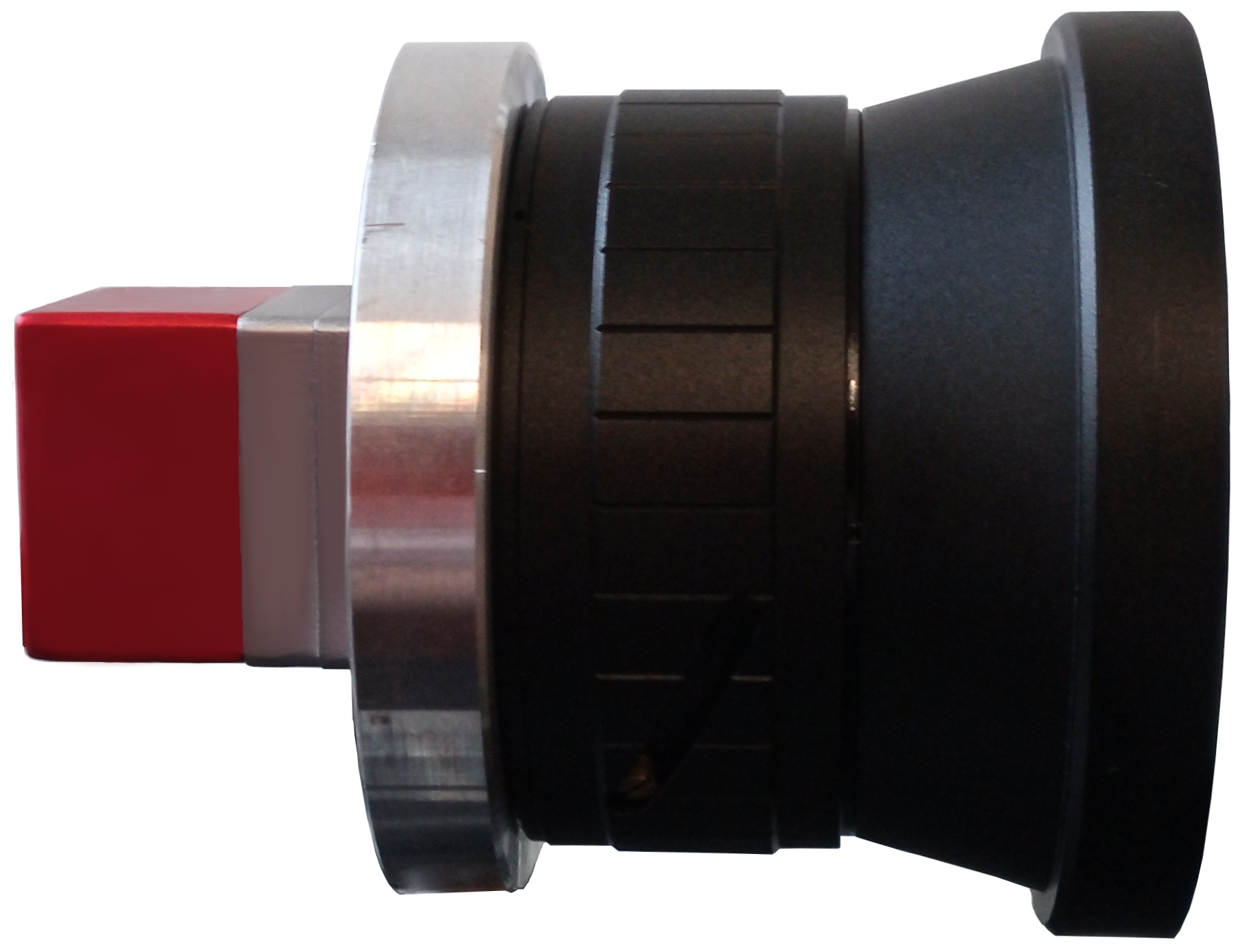}
			};

					\coordinate(lensAssemblyLabelPos) at (0.125,0.69);
					\draw[<-,>=stealth,thick](lensAssemblyLabelPos)--++(0,0.1)node[above,align=center]{camera\\lens};
					\draw[<-,>=stealth,thick](-0.0125,0.605)--++(0,0.0875)node[above,align=center]{camera\\module};
					\node[align=center](adapterRingLabel) at (0.0475,0.1455) {adapter\\plate};
					\draw[->,>=stealth,thick](adapterRingLabel.north)--++(0,0.0875);

					\draw[->,>=stealth,thick](teflonLabel.south)to[in=170,out=-70](0.625,0.85); 

		\end{scope}
\end{tikzpicture}

%% file: sections/pca-imaging_results_v1.tex
\subsection{THz-TDS beam profiling}
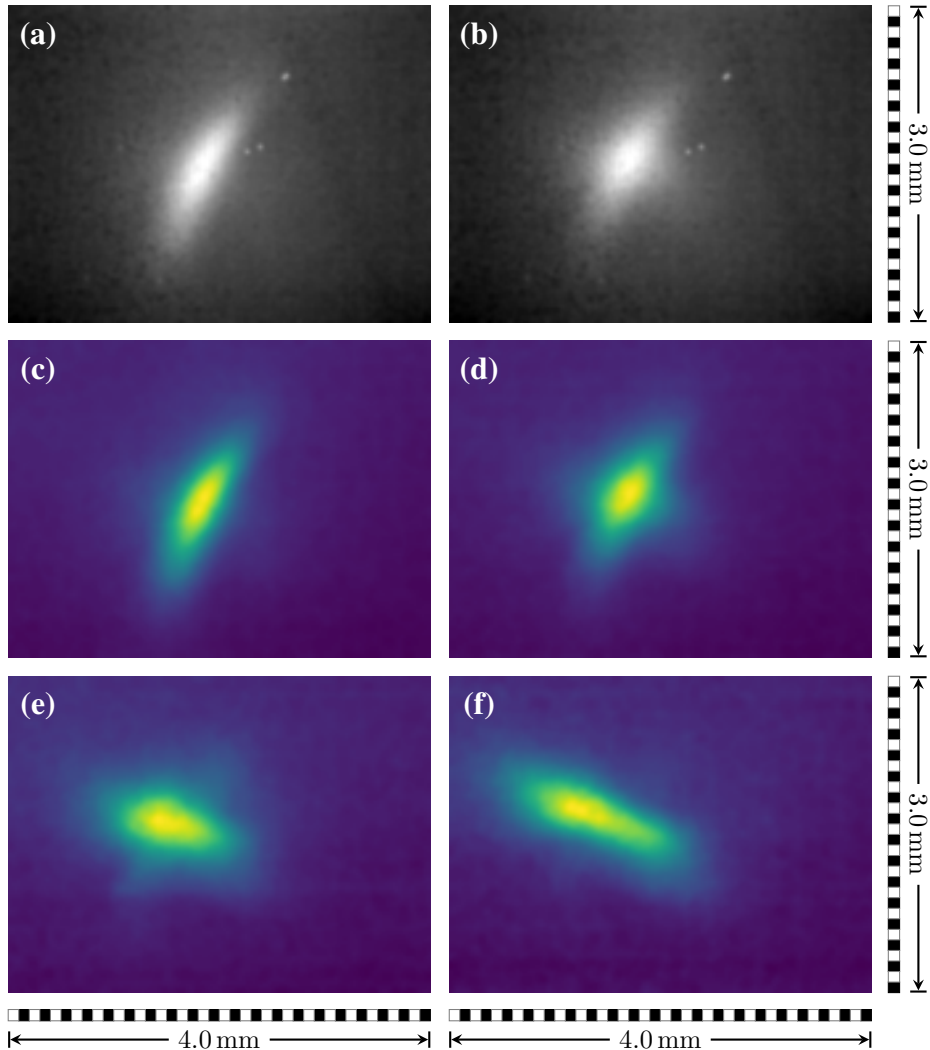
\begin{figure}[h!]
		\centering
		\input{./figures/results/beam-shape/pca-imaging_results_fig3_beam-shape.tikz}
		\caption{
			Selected single frames of a 1D real-time beam shape scan in propagation direction. Data as saved from camera software (a, b), and same data for comparison in false-color with post-processing applied (c, d). Subfigures (a, c, f) depict the beam shape for two out-of-focus positions (before/after focal point). The spatial intensity distribution close to optimal focus is presented in (b, d, e).
		}
		\label{fig:resultsTHzTDSbeamProfiling}
	\end{figure}

		\newcommand{\TXTHzOutputPower}{40}
		\newcommand{\anglePolarizers}{65}
		\newcommand{\assumedLossAlongOpticalPath}{50}
		\pgfmathsetmacro\detectorSize{3*4}
		\pgfmathsetmacro\intensityDropFactor{cos(\anglePolarizers)*cos(\anglePolarizers)}
		\pgfmathsetmacro\intensityDropFactorPercent{100*\intensityDropFactor}
		\pgfmathsetmacro\ThzIntensityPerFullDetectorArea{(\TXTHzOutputPower*\intensityDropFactor*\assumedLossAlongOpticalPath)/(100*\detectorSize)}
		\pgfmathsetmacro\ThzIntensityPerFullDetectorAreaMWPERMSQUARED{round(\ThzIntensityPerFullDetectorArea*10)*100}
	As a first proof-of-concept, the beam profile of the PCA emission was measured with the setup depicted in \mbox{figure \ref{fig:TDSbeamshapeSetupSchematic}}. 
	In this configuration, we have a 1:1 imaging of the beam shape to the sensor.
	Since the intensity of the focused beam was too high for the extremely sensitive camera, the polarizer $\text{P}_2$ was rotated by $\theta\approx\SI{\anglePolarizers}{\degree}$, letting according to Malus's law \mbox{($I=I_0\cdot\cos^2\left(\theta\right)$) \cite{niedrig_optik_2004}} roughly \SI[round-mode=places,round-precision=0]{\intensityDropFactorPercent}{\percent} of the initial intensity pass.
	Assuming \SI{\assumedLossAlongOpticalPath}{\percent} further loss along the optical path, we expect an average intensity of less than \SI[per-mode=symbol,round-mode=places,round-precision=0]{\ThzIntensityPerFullDetectorAreaMWPERMSQUARED}{\milli\watt\per\square\meter} at the detector, but we were still able to obtain decent contrast without any data processing (see figure \ref{fig:resultsTHzTDSbeamProfiling} (a, b)). \newline The images in figure \ref{fig:resultsTHzTDSbeamProfiling} represent single frames of a movie clip (see video S1 in supplementary materials), which was obtained by moving the camera along the THz propagation direction. Data acquisition was performed at \SI{9}{\fps}, allowing the experimenter to get immediate feedback. Even the unprocessed data directly streamed from the camera (\mbox{fig. \ref{fig:resultsTHzTDSbeamProfiling} (a, b)}) provided sufficient information for a qualitative analysis. Post-processing the camera data (\mbox{fig. \ref{fig:resultsTHzTDSbeamProfiling} (c-f)}) revealed that the beam out of focus (\mbox{fig. \ref{fig:resultsTHzTDSbeamProfiling} (c, f)}) was elliptical and tilted about $\pm\SI{45}{\degree}$ to the horizon. Close to the focus (\mbox{fig. \ref{fig:resultsTHzTDSbeamProfiling} (d, e)}), the beam was slightly cross-shaped. Also, the continuous transition from $+\SI{45}{\degree}$ to  $-\SI{45}{\degree}$ tilt could be resolved.

\vspace{-1cm} 
\subsection{Siemens star}
		\newcommand{\SiemensStarOuterArea}{122.702} 
		\newcommand{\SiemensStarInnerArea}{88.438} 
		\pgfmathsetmacro\SiemensStarOuterDiameter{2*sqrt((\SiemensStarOuterArea/3.1415926))} 
		\pgfmathsetmacro\SiemensStarInnerDiameter{2*sqrt((\SiemensStarInnerArea/3.1415926))} 
	
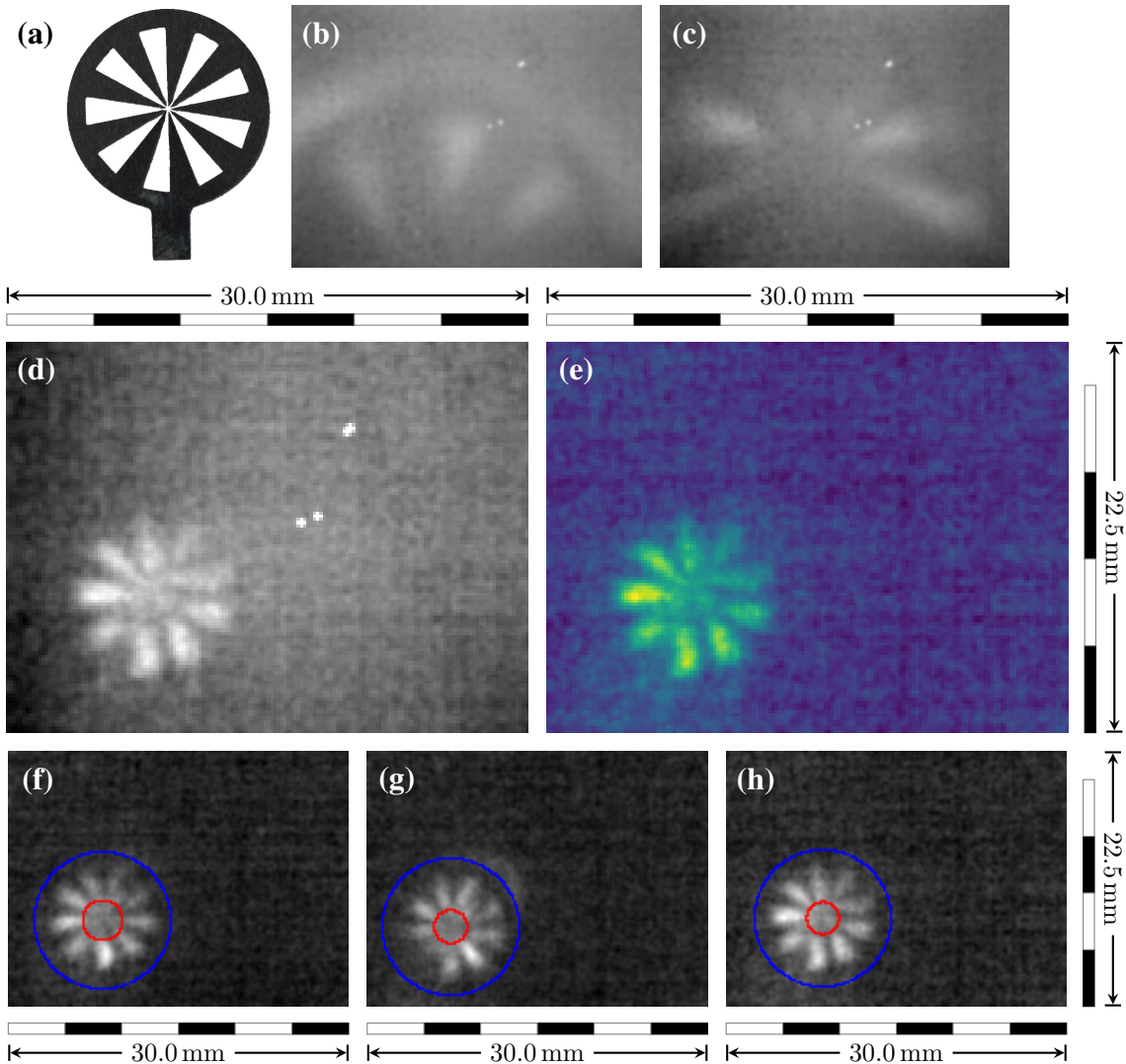
\begin{figure}[h!]
		\centering
		\input{./figures/results/siemens-star/pca-imaging_results_fig4_siemens-star.tikz}
		\caption{
			Imaging of a metallic Siemens star. Photography (VIS) of Siemens star (a).  Unprocessed THz images acquired in zigzag \THzTDS geometry (see fig. \ref{fig:TDSbeamshapeSetupSchematic}): Image with outer rim of Siemens star visible (b), whereas in (c) only central part resolvable. THz images acquired in lens-based setup (see \mbox{fig. \ref{fig:linearImagingSetupSchematic}}):  unprocessed THz data (d) and false-color with dead-pixel-removal applied (e). Determination of resolution shown with three example frames (f-h): Assumed outer rim of the Siemens star (blue circles) and resolution limit (red circles).
		}
		\label{fig:resultsSiemensStar}
	\end{figure}

	The first imaging tests were carried out on a Siemens star (photography with visible light (VIS) in \mbox{fig. \ref{fig:resultsSiemensStar}} (a), outer diameter $d=\SI[round-mode=places,round-precision=1]{\SiemensStarOuterDiameter}{\milli\meter}$, rim diameter $d_{\,\text{rim}}=\SI[round-mode=places,round-precision=1]{\SiemensStarInnerDiameter}{\milli\meter}$, 9 spokes), laser ablated from a thin metal sheet and mounted onto a \SI{1}{\milli\meter} thick sheet of Teflon. 
	To profit from the more intense sample irradiation in the \THzTDS (achievable intensity higher due to smaller beam size), the sample was placed into the standard position (see \mbox{fig. \ref{fig:TDSbeamshapeSetupSchematic}}). By intentionally shifting the first OAPM pair and TX closer to the sample, the focus is moved beyond the sample plane, effectively enlarging the illuminated sample area. With this approach, a part of the Siemens star could be imaged (\mbox{fig. \ref{fig:resultsSiemensStar}} (b)). However, the zigzag configuration with the OAPMs did not allow for an undistorted imaging of such a large sample. This was demonstrated by repositioning the Siemens star only slightly \mbox{(fig. \ref{fig:resultsSiemensStar} (c))}. Switching to the linear setup (\mbox{fig. \ref{fig:linearImagingSetupSchematic}}) allowed to resolve the complete Siemens star \mbox{(fig. \ref{fig:resultsSiemensStar} (d))}. 
	For this data set, we did not use any spatial filtering to avoid its impact on the spatial resolution determination. Only dead pixel removal was applied (\mbox{fig. \ref{fig:resultsSiemensStar} (e)}). The recorded real-time video nicely shows the rotation of the Siemens star (see \mbox{fig. \ref{fig:resultsSiemensStar} (f-h)} and video S2 in supplementary materials), just with some minor intensity fluctuations and shifts. 
	\newline 
	The quality of these images allow an estimation of the spatial resolution. First, the smallest radius $r_{\text{min}}$ of a centered circle is determined for which the average contrast between a spoke and an opening is larger than $\SI{10}{\percent}$ of the highest contrast. 
	The resolution is than $r_{\text{res}}=2\pi\cdot r_{\text{min}}/N$, where $N=9$ is the number of spokes. 
	For the current imaging setup a resolution $r_{\text{res}} = \SI{1.05(15)}{\milli\meter}$ was estimated from ten different Siemens star images. 

\subsection{Key in an envelope}
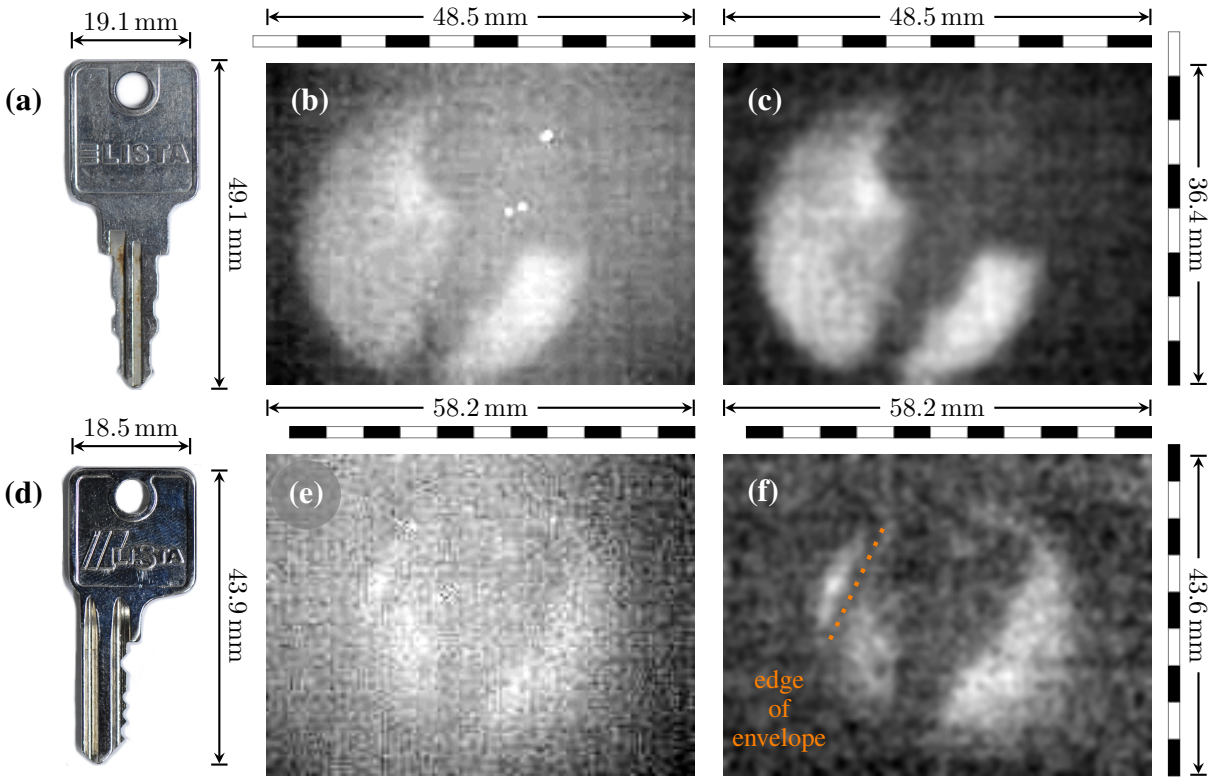
\begin{figure}[h!]
		\centering
		\input{./figures/results/key/pca-imaging_results_fig5_key.tikz}
		\caption{
			Imaging of metal keys. Photography of the keys with rough dimensions (a, d), raw THz images (b, e) of the keys, post-processed 
			THz images (c, f) of the keys with dead pixel removal.  A key can be still resolved within a standard paper envelope (e, f), with edge of envelope marked (f).
		} 
		\label{fig:resultsKey}
	\end{figure}

	We demonstrate the capability of our method for detecting concealed (metallic) objects from a larger distance by examining metallic keys (\mbox{fig. \ref{fig:resultsKey} (a, d)}), which were concealed with a standard paper envelope. Two very similar keys were placed into the THz beam approximately \SI{600}{\milli\meter} away from the camera assembly (see \mbox{fig. \ref{fig:linearImagingSetupSchematic}}). The thermal image was suppressed with \SI{3}{\milli\meter} of Teflon, located between sample and camera. As expected, a metal key without envelope was clearly resolved (\mbox{fig. \ref{fig:resultsKey} (b)}). The post-processing enhanced the contrast and made the edges more defined (\mbox{fig. \ref{fig:resultsKey} (c)}, video S3 in supplementary materials). Putting a metal key into a paper envelope reduced the image quality due to absorption and diffraction on the rough paper surface (\mbox{fig. \ref{fig:resultsKey} (e)}). An additional quality loss originated from saving the data for testing purposes as 8 bit jpg, a format which seems not to be suitable for our THz imaging purposes. Overall, the key shape was fairly faint, but post-processing could enhance the visual clarity so far that even the edge of the paper envelope became visible (\mbox{fig. \ref{fig:resultsKey} (f)}). The video S4 in the supplementary materials shows how the experiment was performed in the laboratory.
	\FloatBarrier

\newpage
\subsection{Leaves with different moisture contents}
\begin{figure}[h!]
		\centering
		\vspace{-0.125cm}
		\input{./figures/results/leaves/pca-imaging_results_fig6_leaves.tikz}
		\caption{
			(a) Photography (VIS) of leaves with different water content. One leaf stored in drawer for five years (top), two picked up just before experiments: One from dry place (middle), one from wet gutter (bottom).
			Glued with THz transparent adhesive tape onto a \SI{1}{\milli\meter} thick Teflon sheet; Teflon and tape stripes removed by image processing for better visual clarity; partially still visible. THz image of the different leaves (b) with a pair of tweezers for contrast enhancement (vertical lines) and the edge of the Teflon sheet (top, horizontal) visible. Higher water contents are clearly represented by decreased brightness. Image stitched with auto-correlation from single frames of a real-time 1D scan. An exemplary post-processed single frame is presented for the dry (c), wet (d) and very wet (e) leaf.
		}
		\label{fig:resultsLeaves}
	\end{figure}
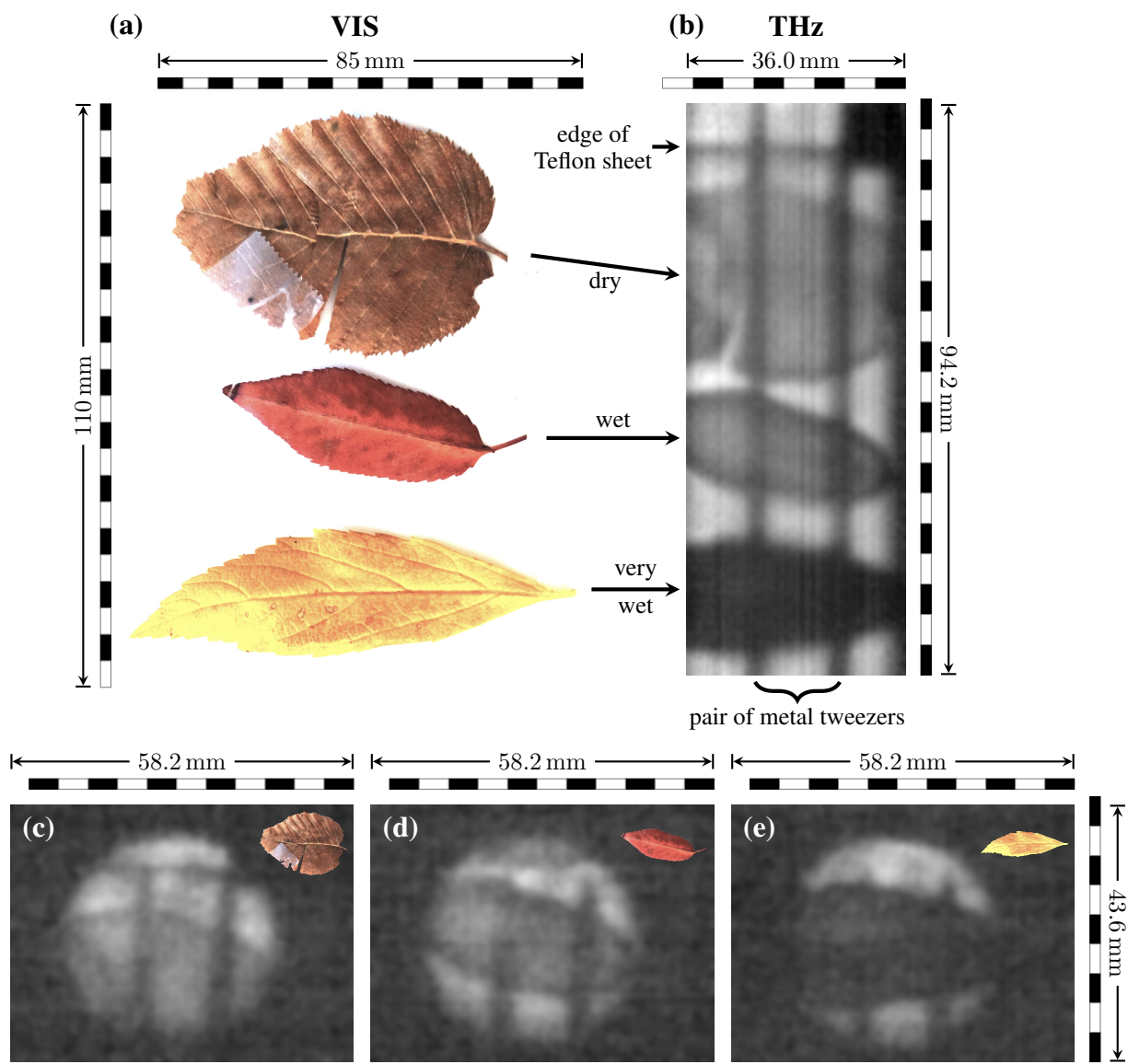

	The strong absorption of water in the THz regime renders THz imaging an interesting modality for biological samples. We evaluated the potential of our approach by investigating leaves with different moisture contents. Three distinct leave specimen (\mbox{fig. \ref{fig:resultsLeaves} (a)}) have been mounted on a \SI{1}{\milli\meter} thick Teflon sheet and were scanned as described in \mbox{subsection \ref{subsec:AnalysisAndPostRoutines} and video S6 in supplementary materials}. The stitched image (\mbox{fig. \ref{fig:resultsLeaves} (b)}) as well as exemplary single frames (\mbox{fig. \ref{fig:resultsLeaves} (c-e)} from video S5 in supplementary materials) provide the same distinct larger features like shape, cracks etc. as the photography (\mbox{fig. \ref{fig:resultsLeaves} (a)}). Additionally, the THz image shows leaves with higher moisture content distinctly darker. 
	Although loosing the ability to resolve finer details, this could allow for an accurate qualitative analysis and even monitoring of diffusion processes in real-time.
	\FloatBarrier
	
\subsection{Thin wood sample}
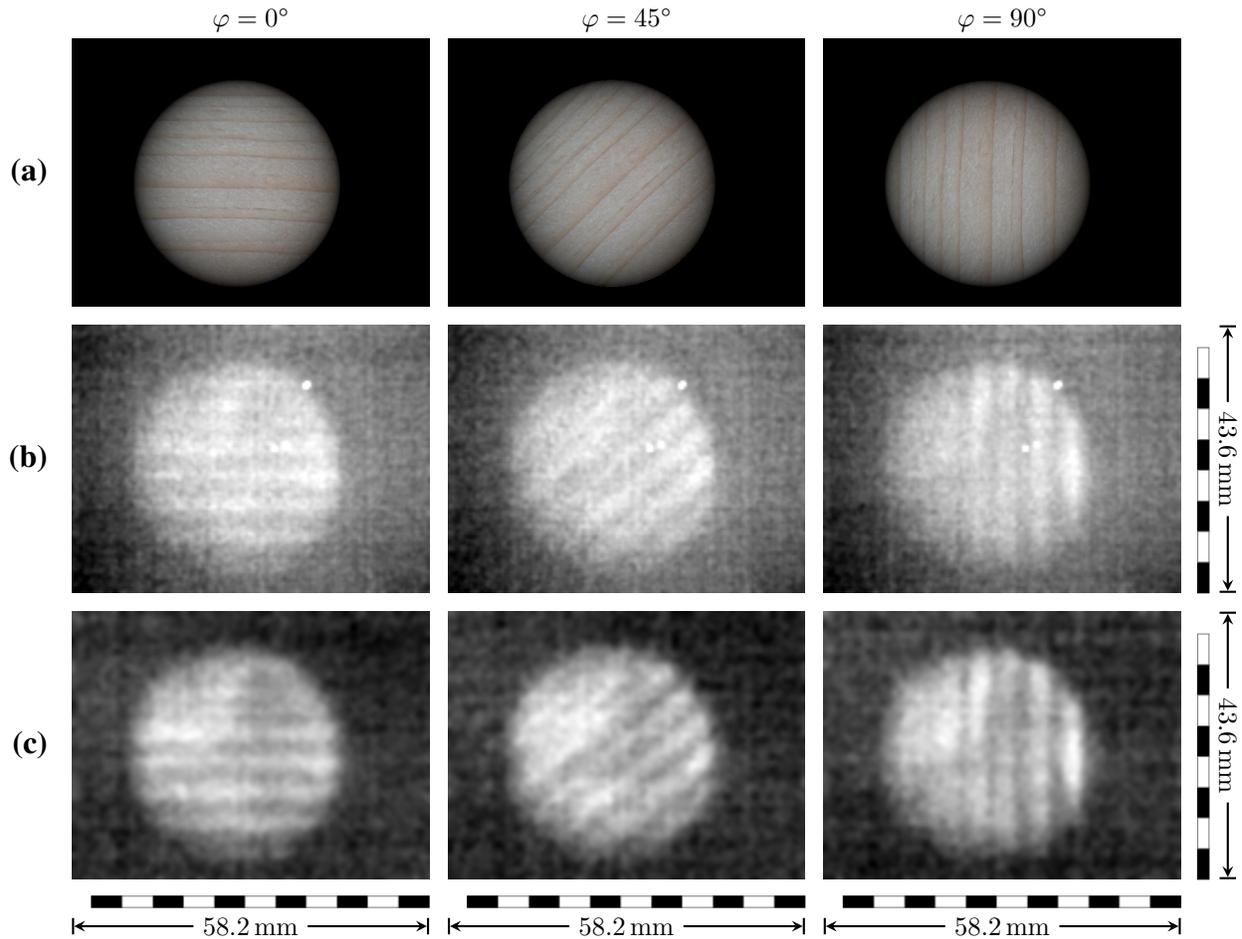
\begin{figure}[h!]
		\centering
		\input{./figures/results/wood/pca-imaging_results_fig7_wood.tikz}
		\caption{
			Imaging of a thin wood sample. (a) Not to scale artistic illustration of the approximate illumination of the 
			thin wood sample for different angles $\varphi$, corresponding to the selected THz single frames from a full sample rotation recorded in real-time (b, c). Unprocessed data (b) as obtained from the camera and post-processed representation with dead pixel removal (c).
		}
		\label{fig:resultsWood}
	\end{figure}

	A $\SI{0.19}{\milli\meter}$ thin, microtome cut wood sample was mounted in a rotatable holder. 
	Thermal image suppression was achieved with \SI{3}{\milli\meter} of Teflon, located between sample and camera. The zero position of the rotation ($\varphi=\SI{0}{\degree}$) was defined such that the annual rings were parallel to the polarization of the THz radiation. This preventive measure allows to determine whether any influence of the polarization on the recorded image exists.
	\newline An approximation for the illuminated area of the actual sample is shown for different orientations in the artistic illustration \mbox{figure \ref{fig:resultsWood} (a)}. The annual rings are already visible in the raw THz images (\mbox{fig. \ref{fig:resultsWood} (b)}) and become more pronounced in the post-processed data (\mbox{fig. \ref{fig:resultsWood} (c)}). The annual rings are clearly recognizable for each configuration. There is no evidence that the ring orientation would influence the image contrast. The THz images \mbox{(fig. \ref{fig:resultsWood} (b, c))} are single frames of a video clip \mbox{(video S7 in supplementary materials)} that we were able to record in real-time despite the high absorption.
	\vfill
	\FloatBarrier

%% file: figures/results/beam-shape/pca-imaging_results_fig3_beam-shape.tikz
\newcommand{\imageWidth}{0.34\linewidth}
\renewcommand{\scalebarDivisionStep}{0.1}

\begin{tikzpicture}
		\coordinate  (topLeft) at (0,0);

				\foreach \image/\imagePos/\imageShiftX/\imageShiftY/\subfig in {%
						240/topLeft/0/0/c,%
						280/beamshape240.north east/0.25/0/d,%
						320/beamshape240.south west/0/-0.25/e,%
						360/beamshape280.south west/0/-0.25/f%
				}{
							\node[anchor=north west,inner sep=0, outer sep=0](beamshape\image) at ($(\imagePos)+(\imageShiftX,\imageShiftY)$){
								\includegraphics[width=\imageWidth]{./figures/results/beam-shape/Beam\image _cm}
					};
					\node[text=white] at ($(beamshape\image.north west)+(4mm,-4mm)$){\large \bfseries (\subfig )};
				}

				\foreach \image/\imagePos/\imageShiftX/\imageShiftY/\subfig in {%
						240/beamshape240.north west/0/0.25/a,%
						280/beamshape240Raw.south east/0.25/0/b%
				}{
					\node[anchor=south west,inner sep=0, outer sep=0](beamshape\image Raw) at ($(\imagePos)+(\imageShiftX,\imageShiftY)$){
								\includegraphics[width=\imageWidth]{./figures/results/beam-shape/Beam\image _raw}
					};
					\node[text=white] at ($(beamshape\image Raw.north west)+(4mm,-4mm)$){\large \bfseries (\subfig )};
				}

			\verticalScalebarRight{beamshape280Raw}{./figures/results/beam-shape/Beam280_raw.csv}{1}{0} 
			\verticalScalebarRight{beamshape280}{./figures/results/beam-shape/Beam280.csv}{1}{0} 
			\verticalScalebarRight{beamshape360}{./figures/results/beam-shape/Beam360.csv}{1}{0} 

			\horizontalScalebarBottom{beamshape360}{./figures/results/beam-shape/Beam360.csv}{1}{0}
			\horizontalScalebarBottom{beamshape320}{./figures/results/beam-shape/Beam320.csv}{1}{0}
\end{tikzpicture}
\renewcommand{\scalebarDivisionStep}{5}

%% file: figures/results/siemens-star/pca-imaging_results_fig4_siemens-star.tikz
\begin{tikzpicture}
				\node[anchor=east,inner sep=0,outer sep=0] (siemensStarTDS1) at (0,0){
					\includegraphics[width=0.285\linewidth]{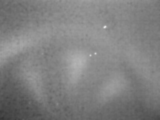}
				};

				\node[anchor=west,inner sep=0,outer sep=0] (siemensStarTDS2) at ($(siemensStarTDS1.east)+(\subfigHorizontalSep,0)$){
					\includegraphics[width=0.285\linewidth]{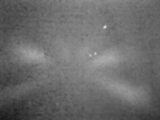}
				};

				\node[anchor=east,inner sep=0,outer sep=0] (siemensStarPhoto) at ($(siemensStarTDS1.west)+(-\subfigHorizontalSep,0)$){
					\includegraphics[height=0.21375\linewidth]{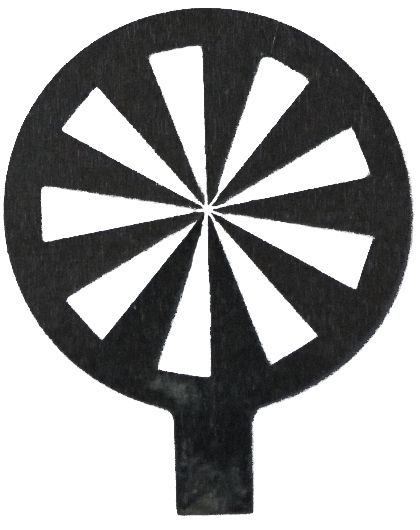}
				};

				\coordinate(toprowCenterBottom) at ($(siemensStarPhoto.south west)!0.5!(siemensStarTDS2.south east)$);
				\node[anchor=north east,inner sep=0,outer sep=0](siemensStarLinearRaw) at ($(toprowCenterBottom)+(-0.5*\subfigHorizontalSep,-\subfigVerticalSep cm -0.75cm)$){
						\includegraphics[width=0.425\linewidth]{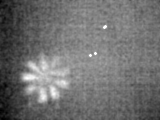}
				};

				\node[anchor=west,inner sep=0,outer sep=0](siemensStarLinearEdit) at ($(siemensStarLinearRaw.east)+(\subfigHorizontalSep,0)$){
					\includegraphics[width=0.425\linewidth]{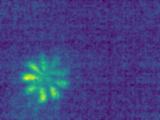}
				};

				\coordinate(centerrowCenterBottom) at ($(siemensStarLinearRaw.south west)!0.5!(siemensStarLinearEdit.south east)$);
				\node[anchor=north,inner sep=0,outer sep=0](siemensStarRes2) at ($(centerrowCenterBottom)+(0,-\subfigVerticalSep)$){
					\includegraphics[width=0.2775\linewidth]{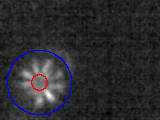}
				};

				\node[anchor=west,inner sep=0,outer sep=0](siemensStarRes3) at ($(siemensStarRes2.east)+(\subfigHorizontalSep,0)$){
					\includegraphics[width=0.2775\linewidth]{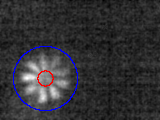}
				};

				\node[anchor=east,inner sep=0,outer sep=0](siemensStarRes1) at ($(siemensStarRes2.west)+(-\subfigHorizontalSep,0)$){
					\includegraphics[width=0.2775\linewidth]{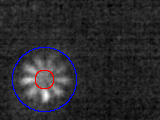}
				};

								\horizontalScalebarTop{siemensStarLinearRaw}{./figures/results/siemens-star/0406_raw.csv}{1}{0}
								\horizontalScalebarTop{siemensStarLinearEdit}{./figures/results/siemens-star/0406.csv}{1}{0}
								\verticalScalebarRight{siemensStarLinearEdit}{./figures/results/siemens-star/0406.csv}{1}{0}
								\horizontalScalebarBottom{siemensStarRes1}{./figures/results/siemens-star/0181_resolution.csv}{1}{0}
								\horizontalScalebarBottom{siemensStarRes2}{./figures/results/siemens-star/0271_resolution.csv}{1}{0}
								\horizontalScalebarBottom{siemensStarRes3}{./figures/results/siemens-star/0406_resolution.csv}{1}{0}
								\verticalScalebarRight{siemensStarRes3}{./figures/results/siemens-star/0406_resolution.csv}{1}{0}

				\foreach \subfigCoord/\subfigLabel/\subfigLabelColor in {%
						siemensStarTDS1/b/white,%
						siemensStarTDS2/c/white,%
						siemensStarLinearRaw/d/white,%
						siemensStarLinearEdit/e/white,%
						siemensStarRes1/f/white,%
						siemensStarRes2/g/white,%
						siemensStarRes3/h/white%
				}{
						\node[text=\subfigLabelColor] at ($(\subfigCoord.north west)+(4mm,-4mm)$){\large \bfseries (\subfigLabel )};
				}

				\node[text=black] at ($(siemensStarPhoto.north west)+(-4mm,-4mm)$){\large \bfseries (a)};
\end{tikzpicture}

%% file: figures/results/key/pca-imaging_results_fig5_key.tikz
\begin{tikzpicture}
				\node[anchor=north east,inner sep=0, outer sep=0] (keyPhoto) at (0,0){
						\includegraphics[width=18.5mm]{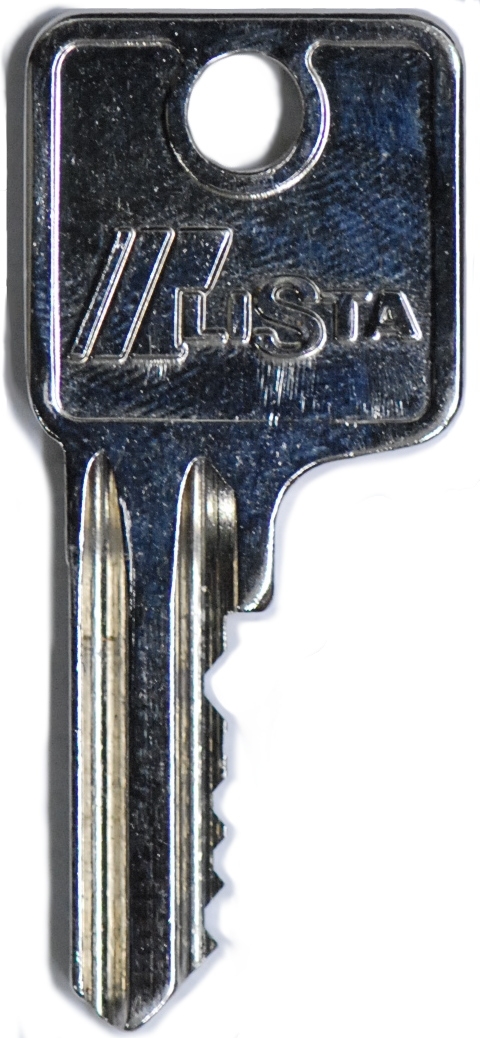}
					};
				\node[anchor=south east,inner sep=0, outer sep=0] (firstKeyPhoto) at ($(keyPhoto.north east)+(0, 1cm)$){
					\includegraphics[width=18.5mm]{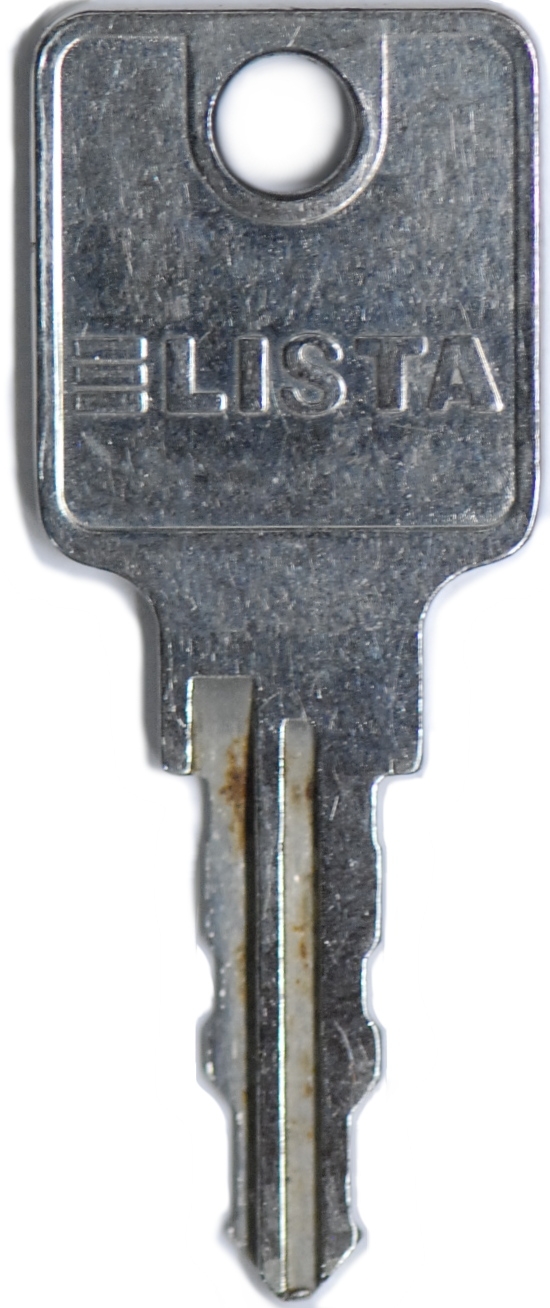}
				};

				\node[anchor=west,inner sep=0, outer sep=0](keyTHz) at ($(firstKeyPhoto.east)+(0.875 cm,0)$){
					\includegraphics[width=0.345\linewidth]{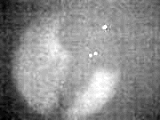}
				};
				\node[anchor=west,inner sep=0, outer sep=0](keyTHzEDIT) at ($(keyTHz.east)+(0.375 cm,0)$){
					\includegraphics[width=0.345\linewidth]{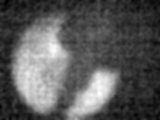}
				};

				\node[anchor=west,inner sep=0, outer sep=0](keyTHzEnvelope) at ($(keyPhoto.east)+(0.875 cm,0)$){
					\includegraphics[width=0.345\linewidth]{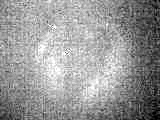}
				};
				\node[anchor=west,inner sep=0, outer sep=0](keyTHzEnvelopeEDIT) at ($(keyTHzEnvelope.east)+(0.375 cm,0)$){
					\includegraphics[width=0.345\linewidth]{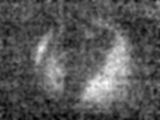}
				};

								\coordinate(startEnvelopBottomY) at ($(keyTHzEnvelopeEDIT.south west)!0.425!(keyTHzEnvelopeEDIT.north west)$);
								\coordinate(startEnvelopBottomX) at ($(keyTHzEnvelopeEDIT.south west)!0.25!(keyTHzEnvelopeEDIT.south east)$);
								\path let \p1=(startEnvelopBottomX), \p2=(startEnvelopBottomY) in (\x1,\y2)coordinate(startEnvelopBottom);

								\coordinate(startEnvelopTopY) at ($(keyTHzEnvelopeEDIT.south west)!0.775!(keyTHzEnvelopeEDIT.north west)$);
								\coordinate(startEnvelopTopX) at ($(keyTHzEnvelopeEDIT.south west)!0.375!(keyTHzEnvelopeEDIT.south east)$);
								\path let \p1=(startEnvelopTopX), \p2=(startEnvelopTopY) in (\x1,\y2)coordinate(startEnvelopTop);

								\draw[orange,loosely dotted,ultra thick]++(startEnvelopBottom)--(startEnvelopTop);
								\node[text=orange,anchor=south west,align=center](envEdgeLabel) at ($(keyTHzEnvelopeEDIT.south west)+(0,2mm)$){edge\\ of\\ envelope};

					\draw[|<->|,thick,>=stealth]($(firstKeyPhoto.south east)+(0.25,0)$)--($(firstKeyPhoto.north east)+(0.25,0)$);
					\node[anchor=center,rotate=-90] at ($(firstKeyPhoto.east)+(0.5,0)$){\SI{49.1}{\milli\metre}};
					\draw[|<->|,thick,>=stealth]($(firstKeyPhoto.north west)!0.0675!(firstKeyPhoto.north east)+(0,0.25)$)--($(firstKeyPhoto.north east)!0.0675!(firstKeyPhoto.north west)+(0,0.25)$);
					\node[anchor=center] at ($(firstKeyPhoto.north)+(0,0.5)$){\SI{19.1}{\milli\metre}};

					\draw[|<->|,thick,>=stealth]($(keyPhoto.south east)+(0.25,0)$)--($(keyPhoto.north east)!0.015!(keyPhoto.south east)+(0.25,0)$);
					\node[anchor=center,rotate=-90] at ($(keyPhoto.east)+(0.5,0)$){\SI{43.9}{\milli\metre}};
					\draw[|<->|,thick,>=stealth]($(keyPhoto.north west)!0.0725!(keyPhoto.north east)$)++(0,0.25)--($(keyPhoto.north east)!0.06725!(keyPhoto.north west)+(0,0.25)$);
					\node[anchor=center] at ($(keyPhoto.north)+(0,0.5)$){\SI{18.5}{\milli\metre}};

					\horizontalScalebarTop{keyTHz}{./figures/results/key/Key10_raw.csv}{1}{0} 
					\horizontalScalebarTop{keyTHzEDIT}{./figures/results/key/Key10.csv}{1}{0} 
					\horizontalScalebarTop{keyTHzEnvelope}{./figures/results/key/KeyInEnvelop19_raw.csv}{1}{1} 
					\horizontalScalebarTop{keyTHzEnvelopeEDIT}{./figures/results/key/KeyInEnvelop19.csv}{1}{1} 

					\verticalScalebarRight{keyTHzEDIT}{./figures/results/key/Key10.csv}{1}{0} 
					\verticalScalebarRight{keyTHzEnvelopeEDIT}{./figures/results/key/KeyInEnvelop19.csv}{0}{0} 

						\node[anchor=north west, font=\large,text=white](labelB) at ($(keyTHz.north west)+(2mm,-2mm)$){\bfseries (b)};
						\node[anchor=north west, font=\large,text=white](labelC) at ($(keyTHzEDIT.north west)+(2mm,-2mm)$){\bfseries (c)};
						\path let \p1=($(firstKeyPhoto.north west)+(-5mm,0)$), \p2=(labelB) in (\x1,\y2) node[font=\large]{\bfseries (a)};

						\node[anchor=north west, font=\large,text=white,opacity=.675,text opacity=1,fill=gray,circle](labelE) at ($(keyTHzEnvelope.north west)+(2mm,-2mm)$){\bfseries(e)};
						\node[anchor=north west, font=\large,text=white](labelF) at ($(keyTHzEnvelopeEDIT.north west)+(2mm,-2mm)$){\bfseries (f)};
						\path let \p1=($(firstKeyPhoto.north west)+(-5mm,0)$), \p2=(labelE) in (\x1,\y2) node[font=\large]{\bfseries (d)};

\end{tikzpicture}

%% file: figures/results/leaves/pca-imaging_results_fig6_leaves.tikz
\begin{tikzpicture}
				\node[inner sep=0, outer sep=0](leavesVIS) at (0,0){
					\includegraphics[height=0.5\linewidth]{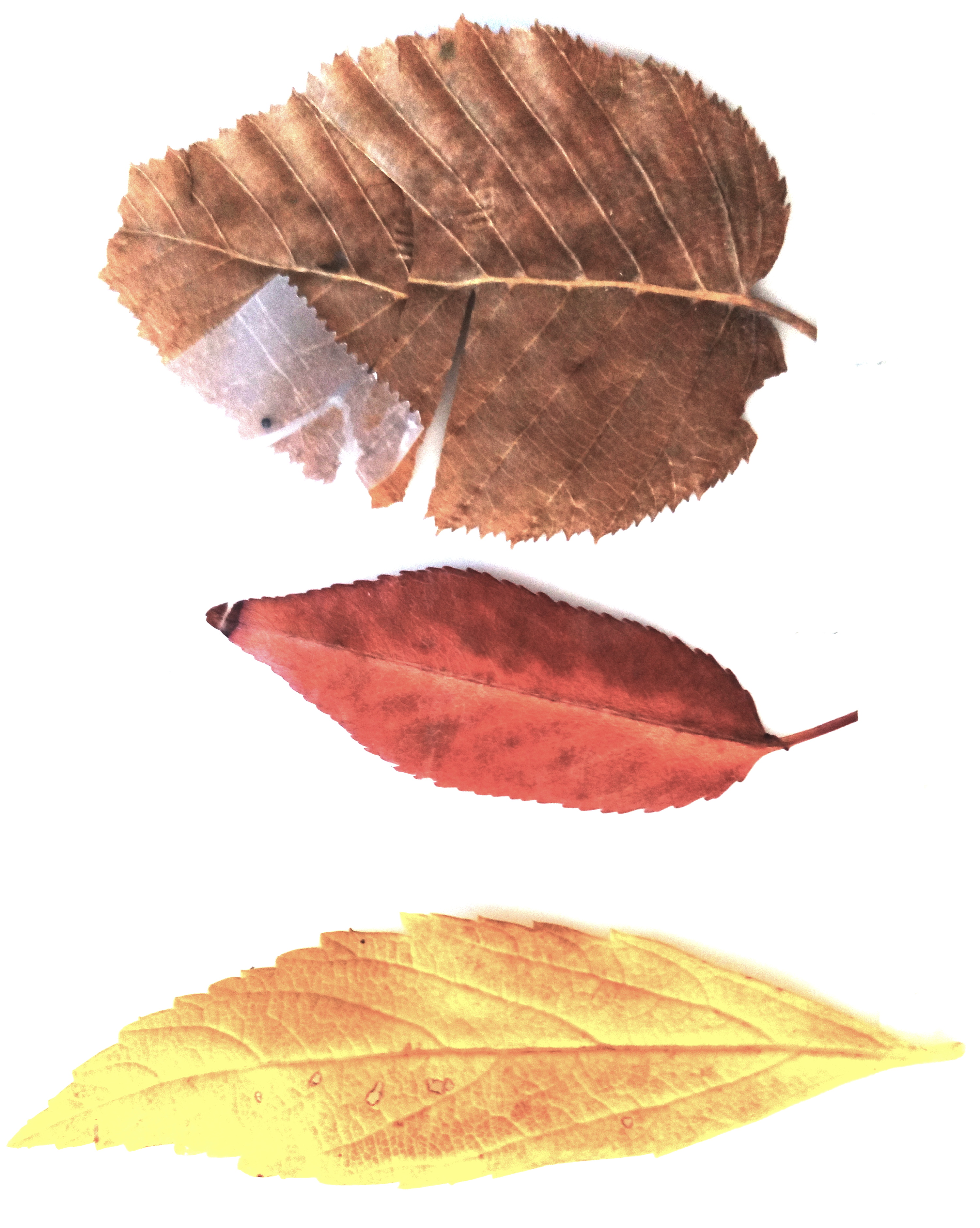}
				};
				\node[anchor=south west,inner sep=0, outer sep=0](leavesStitched) at ($(leavesVIS.south east)+(1.5 cm,0cm)$){
					\includegraphics[height=0.5\linewidth]{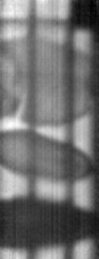}
				};

				\node[anchor=north west,inner sep=0, outer sep=0](leavesTHzRawDRY) at ($(leavesVIS.south west)+(-1.675 cm,-1.875cm)$){
					\includegraphics[width=0.3\linewidth]{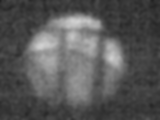}
				};
				\node[anchor=west,inner sep=0, outer sep=0](leavesTHzRawWET) at ($(leavesTHzRawDRY.east)+(0.25 cm,0cm)$){
					\includegraphics[width=0.3\linewidth]{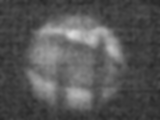}
				};
				\node[anchor=west,inner sep=0, outer sep=0](leavesTHzRawVERYWET) at ($(leavesTHzRawWET.east)+(0.25 cm,0cm)$){
					\includegraphics[width=0.3\linewidth]{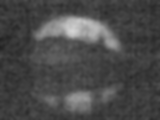}
				};

				\node[anchor=north east,inner sep=0, outer sep=0](dryThumbnail) at ($(leavesTHzRawDRY.north east)+(-1mm,-1mm)$){
					\includegraphics[width=1.25cm]{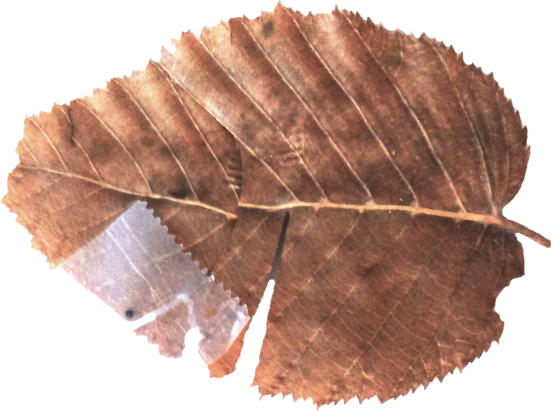}
				};
				\path let \p1=(leavesTHzRawWET.north east), \p2=(dryThumbnail.center) in (\x1-1mm,\y2) node[anchor=east,inner sep=0, outer sep=0]{
					\includegraphics[width=1.25cm]{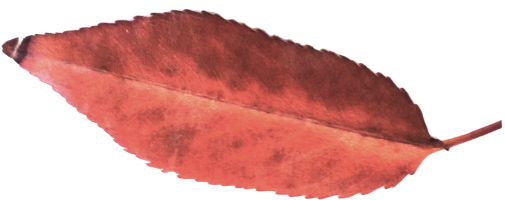}
				};
				\path let \p1=(leavesTHzRawVERYWET.north east), \p2=(dryThumbnail.center) in (\x1-1mm,\y2) node[anchor=east,inner sep=0, outer sep=0]{
					\includegraphics[width=1.25cm]{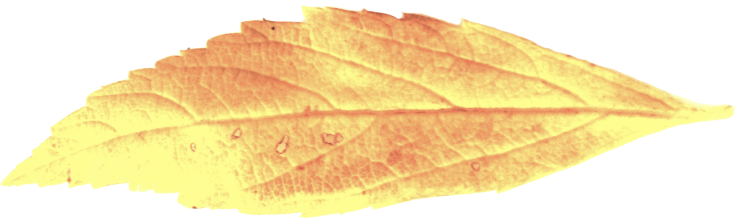}
				};

				\path ($(leavesVIS.north east)!0.265!(leavesVIS.south east)$)--++(-0.75cm,0)coordinate(dryVIS);
				\path ($(leavesVIS.north east)!0.585!(leavesVIS.south east)$)--++(-0.525cm,0)coordinate(wetVIS);
				\path ($(leavesVIS.north east)!0.85!(leavesVIS.south east)$)--++(0.125 cm,0)coordinate(wetterVIS);

				\path ($(leavesVIS.north east)!0.265!(leavesVIS.south east)$)--++(-1.535cm,0.175cm)coordinate(dryVIS_SCALEEAST);
				\path ($(leavesVIS.north east)!0.265!(leavesVIS.south east)$)--++(-1.25cm-4.1875cm,6.25mm)coordinate(dryVIS_SCALEWEST);
				\path ($(leavesVIS.north east)!0.5!(leavesVIS.north west)$)--++(0.1207535cm,-0.25cm)coordinate(dryVIS_SCALENORTH);
				\path ($(leavesVIS.north east)!0.5!(leavesVIS.south east)$)--++(-3.445cm,6.5mm)coordinate(dryVIS_SCALESOUTH);

				\path ($(leavesStitched.north west)!0.3!(leavesStitched.south west)$)--++(-0.1cm,0)coordinate(dryTHZ);
				\path ($(leavesStitched.north west)!0.585!(leavesStitched.south west)$)--++(-0.1cm,0)coordinate(wetTHZ);
				\path ($(leavesStitched.north west)!0.85!(leavesStitched.south west)$)--++(-0.1 cm,0)coordinate(wetterTHZ);

				\path ($(leavesStitched.north west)!0.075!(leavesStitched.south west)$)--++(-0.1cm,0)coordinate(teflonEdgeTopLeft);

				\path ($(leavesStitched.south)+(0,-0.1cm)$)coordinate(leavesStitchedTweezersLabelBottom);

				\path ($(leavesStitched.south west)!0.3125!(leavesStitched.south east)$)--++(0,-0.1cm)coordinate(leavesStitchedTweezersLeftBottom);
				\path ($(leavesStitched.south west)!0.7!(leavesStitched.south east)$)--++(0,-0.1cm)coordinate(leavesStitchedTweezersRightBottom);

%
%
%

									\newcommand{\distanceImageJ}{54.451} 
									\newcommand{\desiredScalebarLength}{85} 
									\pgfmathsetmacro\scalebarLengthFactor{\desiredScalebarLength/\distanceImageJ}
									\pgfmathsetmacro\scalebarNumberOfSteps{\desiredScalebarLength/\scalebarDivisionStep}
									\pgfmathsetmacro\scalebarNumberOfStepsBlack{\scalebarNumberOfSteps-1}
									\pgfmathsetmacro\scalebarPercentSteps{1/\scalebarNumberOfSteps}
									\coordinate (scalebarBottomStartEast) at ($(leavesVIS.north east)+(0,0.375)$);

									\path[draw=black,line width=\scalebarHeight mm] let \p1 = (dryVIS_SCALEWEST), \p2=(dryVIS_SCALEEAST), \n1={{veclen(\x2-\x1,\y2-\y1)}} in ($(leavesVIS.north east)+(0,0.3)$)--++(180:{\scalebarLengthFactor*\n1})node[anchor=east](endOfScalebar){}\pgfextra{\xdef\var{\n1}} ;

									\path let \p1=(endOfScalebar.east), \p2=(scalebarBottomStartEast) in (\x1,\y2) coordinate(scalebarMover);

									\foreach \x in {0,2,...,\scalebarNumberOfStepsBlack}{
											\pgfmathsetmacro\percentHelperStart{\x*\scalebarPercentSteps}
											\pgfmathsetmacro\percentHelper{(\x+1)*\scalebarPercentSteps}
											\coordinate(rectangleStart) at ($(scalebarBottomStartEast)!\percentHelperStart!(scalebarMover)$);
											\coordinate (scalebarRectangleHelper) at ($(scalebarBottomStartEast)!\percentHelper!(scalebarMover)$);
											\coordinate(rectangleEnd) at ($(scalebarRectangleHelper)+(0,-\scalebarHeight mm)$);
											\fill[black,draw=black](rectangleStart)rectangle(rectangleEnd);
											\coordinate(blackLabel\x) at ($(rectangleStart)!0.5!(scalebarRectangleHelper)$);
									}

									\pgfmathsetmacro\noWhiteParts{\scalebarNumberOfSteps-1}
									\foreach \x in {1,3,...,\noWhiteParts}{
											\pgfmathsetmacro\percentHelperStart{\x*\scalebarPercentSteps}
											\pgfmathsetmacro\percentHelper{(\x+1)*\scalebarPercentSteps}
											\coordinate (scalebarRectangleHelper) at ($(scalebarBottomStartEast)!\percentHelper!(scalebarMover)$);
											\fill[white,draw=gray]($(scalebarBottomStartEast)!\percentHelperStart!(scalebarMover)$)rectangle($(scalebarRectangleHelper)+(0,-\scalebarHeight mm)$);
									}

									\path ($(scalebarBottomStartEast)!0.5!(scalebarMover)$)++(0,0.25)node(visHorLengthLabel){\SI{85}{\milli\meter}};
									\draw[->|,thick,>=stealth](visHorLengthLabel.east)--($(scalebarBottomStartEast)+(0,0.25)$);
									\draw[->|,thick,>=stealth](visHorLengthLabel.west)--($(scalebarMover)+(0,0.25)$);

									\newcommand{\distanceImageJVertical}{42.233} 
									\newcommand{\desiredScalebarLengthVertical}{110} 
									\pgfmathsetmacro\scalebarLengthFactor{\desiredScalebarLengthVertical/\distanceImageJVertical}
									\pgfmathsetmacro\scalebarNumberOfSteps{\desiredScalebarLengthVertical/\scalebarDivisionStep}
									\pgfmathsetmacro\scalebarNumberOfStepsBlack{\scalebarNumberOfSteps-1}
									\pgfmathsetmacro\scalebarPercentSteps{1/\scalebarNumberOfSteps}
									\coordinate (scalebarLeftStartNorth) at ($(leavesVIS.north west)+(-0.375,0)$);

									\path let \p1 = (dryVIS_SCALESOUTH), \p2=(dryVIS_SCALENORTH), \n1={{veclen(\x2-\x1,\y2-\y1)}} in ($(leavesVIS.north west)+(-0.3,0)$)--++(-90:{\scalebarLengthFactor*\n1})node[anchor=south](endOfScalebar){}\pgfextra{\xdef\var{\n1}};
									\path let \p1=(endOfScalebar.south), \p2=(scalebarLeftStartNorth) in (\x2,\y1) coordinate(scalebarMover);

									\foreach \x in {0,2,...,\scalebarNumberOfStepsBlack}{
											\pgfmathsetmacro\percentHelperStart{\x*\scalebarPercentSteps}
											\pgfmathsetmacro\percentHelper{(\x+1)*\scalebarPercentSteps}
											\coordinate(rectangleStart) at ($(scalebarLeftStartNorth)!\percentHelperStart!(scalebarMover)$);
											\coordinate (scalebarRectangleHelper) at ($(scalebarLeftStartNorth)!\percentHelper!(scalebarMover)$);
											\coordinate(rectangleEnd) at ($(scalebarRectangleHelper)+(\scalebarHeight mm,0)$);
											\fill[black,draw=black](rectangleStart)rectangle(rectangleEnd);
											\coordinate(blackLabel\x) at ($(rectangleStart)!0.5!(scalebarRectangleHelper)$);
									}

									\pgfmathsetmacro\noWhiteParts{\scalebarNumberOfSteps-1}
									\foreach \x in {1,3,...,\noWhiteParts}{
											\pgfmathsetmacro\percentHelperStart{\x*\scalebarPercentSteps}
											\pgfmathsetmacro\percentHelper{(\x+1)*\scalebarPercentSteps}
											\coordinate (scalebarRectangleHelper) at ($(scalebarLeftStartNorth)!\percentHelper!(scalebarMover)$);
											\fill[white,draw=gray]($(scalebarLeftStartNorth)!\percentHelperStart!(scalebarMover)$)rectangle($(scalebarRectangleHelper)+(\scalebarHeight mm,0)$);
									}

									\path ($(scalebarLeftStartNorth)!0.5!(scalebarMover)$)++(-0.25,0)node[rotate=90](visVerLengthLabel){\SI{\desiredScalebarLengthVertical}{\milli\meter}};
									\draw[->|,thick,>=stealth](visVerLengthLabel.east)--($(scalebarLeftStartNorth)+(-0.25,0)$);
									\draw[->|,thick,>=stealth](visVerLengthLabel.west)--($(scalebarMover)+(-0.25,0)$);

						\horizontalScalebarTop{leavesStitched}{./figures/results/leaves/LeavesSwissTHzStitchedImage1.csv}{0}{0} 
						\verticalScalebarRight{leavesStitched}{./figures/results/leaves/LeavesSwissTHzStitchedImage1.csv}{0}{0} 

						\horizontalScalebarTop{leavesTHzRawDRY}{./figures/results/leaves/Leaves32.csv}{0}{1} 
						\horizontalScalebarTop{leavesTHzRawWET}{./figures/results/leaves/Leaves106.csv}{0}{1} 
						\horizontalScalebarTop{leavesTHzRawVERYWET}{./figures/results/leaves/Leaves153.csv}{0}{1} 
						\verticalScalebarRight{leavesTHzRawVERYWET}{./figures/results/leaves/Leaves153.csv}{0}{0} 

				\foreach \coord in {dry,wet,wetter}{
					\draw[->,>=stealth, ultra thick](\coord VIS)--(\coord THZ);
					\coordinate (\coord ArrowCenter) at ($(\coord VIS)!0.5!(\coord THZ)$);
				}

				\node at ($(dryArrowCenter)+(0,-0.25cm)$){dry};
				\node at ($(wetArrowCenter)+(0,0.25cm)$){wet};
				\node at ($(wetterArrowCenter)+(0,0.25cm)$){very};
				\node at ($(wetterArrowCenter)+(0,-0.25cm)$){wet};

				\node[align=center,inner sep=0, outer sep=0](teflonEdgeLabel) at ($(teflonEdgeTopLeft)+(-1.25cm,0)$){edge of\\Teflon sheet};
				\draw[->,>=stealth, ultra thick](teflonEdgeLabel.east)--(teflonEdgeTopLeft);

				\draw[black,ultra thick,decorate,decoration={brace,amplitude=7.5pt,mirror,raise=0pt}](leavesStitchedTweezersLeftBottom)--(leavesStitchedTweezersRightBottom)node[midway,yshift=-15pt]{pair of metal tweezers};

					\node at ($(leavesVIS.north)+(0,1.125cm)$){\large \bfseries VIS};
				\node at ($(leavesStitched.north)+(0,1.125cm)$){\large \bfseries THz};

				\node at ($(leavesVIS.north west)+(0,1.125cm)$){\large \bfseries (a)};
				\node at ($(leavesStitched.north west)+(0,1.125cm)$){\large \bfseries (b)};

				\foreach \subfig/\subfigLabel in {%
					leavesTHzRawDRY/c,%
					leavesTHzRawWET/d,%
					leavesTHzRawVERYWET/e%
				}{
						\node[anchor=north west, font=\large,text=white](subfigLabel\subfig) at ($(\subfig.north west)+(0.5mm,-0.5mm)$){\bfseries (\subfigLabel)};
				}
\end{tikzpicture}

%% file: figures/results/wood/pca-imaging_results_fig7_wood.tikz
\newcommand{\vignette}[4]{
	\draw[fill, even odd rule] (#1.south west) rectangle (#1.north east) (#1.center)++(#2,#3) circle (#4);
}

\begin{tikzpicture}
				\node[anchor=north,inner sep=0, outer sep=0] (wood0) at (0,0){
					\includegraphics[width=0.2875\linewidth]{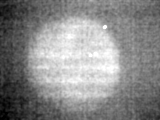}
				};
				\node[anchor=west,inner sep=0, outer sep=0] (wood45) at ($(wood0.east)+(0.25,0)$){
					\includegraphics[width=0.2875\linewidth]{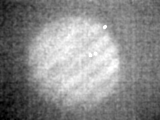}
				};

				\node[anchor=west,inner sep=0, outer sep=0] (wood90) at ($(wood45.east)+(0.25,0)$){
					\includegraphics[width=0.2875\linewidth]{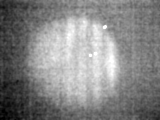}
				};

				\node[anchor=north,inner sep=0, outer sep=0] (woodFCP0) at ($(wood0.south)+(0,-\subfigVerticalSep cm)$){
					\includegraphics[width=0.2875\linewidth]{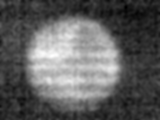}
				};
				\node[anchor=west,inner sep=0, outer sep=0] (woodFCP45) at ($(woodFCP0.east)+(0.25,0)$){
					\includegraphics[width=0.2875\linewidth]{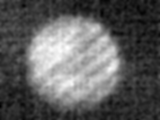}
				};

				\node[anchor=west,inner sep=0, outer sep=0] (woodFCP90) at ($(woodFCP45.east)+(0.25,0)$){
					\includegraphics[width=0.2875\linewidth]{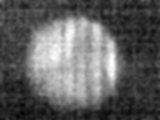}
				};

				\node[anchor=south west,inner sep=0, outer sep=0](woodVIS0) at ($(wood0.north west)+(0,\subfigVerticalSep cm)$){
					\includegraphics[width=0.2875\linewidth]{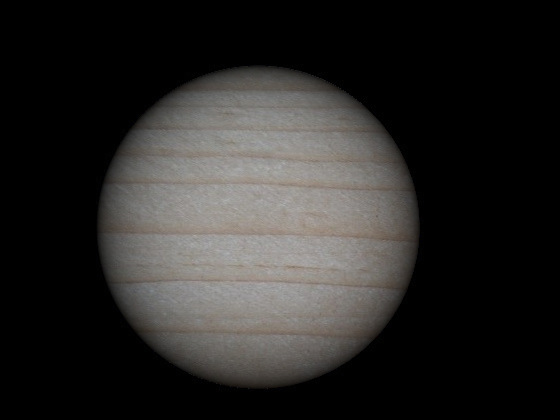}
				};
				\node[anchor=west,inner sep=0, outer sep=0](woodVIS45) at ($(woodVIS0.east)+(0.25,0)$){
					\includegraphics[width=0.2875\linewidth]{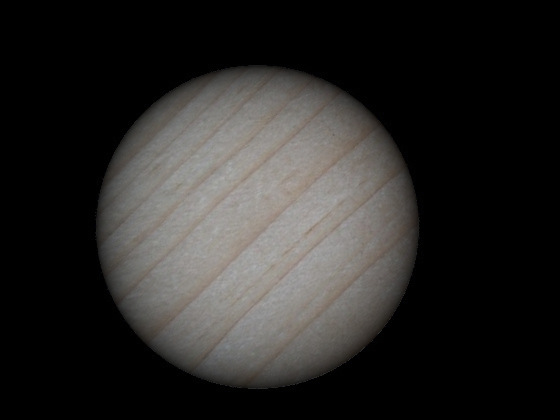}
				};
				\node[anchor=west,inner sep=0, outer sep=0](woodVIS90) at ($(woodVIS45.east)+(0.25,0)$){
					\includegraphics[width=0.2875\linewidth]{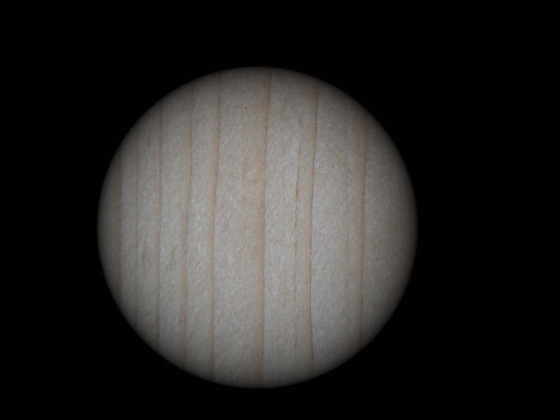}
				};

			\node at ($(woodVIS0.north)+(0,0.25)$){$\varphi=\SI{0}{\degree}$};
			\node at ($(woodVIS45.north)+(0,0.25)$){$\varphi=\SI{45}{\degree}$};
			\node at ($(woodVIS90.north)+(0,0.25)$){$\varphi=\SI{90}{\degree}$};

			\verticalScalebarRight{wood90}{./figures/results/wood/Wood218_raw.csv}{1}{0} 
			\verticalScalebarRight{woodFCP90}{./figures/results/wood/Wood218.csv}{1}{0} 
			\horizontalScalebarBottom{woodFCP90}{./figures/results/wood/Wood218.csv}{0}{1}
			\horizontalScalebarBottom{woodFCP45}{./figures/results/wood/Wood131.csv}{0}{1}
			\horizontalScalebarBottom{woodFCP0}{./figures/results/wood/Wood40.csv}{0}{1}

			\foreach \subfigCoord/\subfigLabel/\subfigLabelColor in {%
					woodVIS0/a/black,%
					wood0/b/black,%
					woodFCP0/c/black%
			}{
					\node[text=\subfigLabelColor, anchor= east] at ($(\subfigCoord.west)+(-2mm,0)$){\large \bfseries (\subfigLabel )};
			}

\end{tikzpicture}

%% file: sections/pca-imaging_discussion_v1.tex
\subsection{Limitations and potential improvements of current setup}
	The major limitation of using a PCA as a THz source is the low output power. This is less relevant as long as images are acquired with a focused beam. However, for a collimated beam, the irradiance decreases quadratically with the beam radius. For example, a strong transmission signal through plastic (food) containers could be detected in the THz-TDS (focused, RX), but with the collimated beam and camera instead of RX, no valid signal could be recorded. Although the camera is extremely sensitive, the expanded beam combined with the sample absorption did not provide enough irradiance at the sensor for real-time imaging. For the current imaging setup, a resolution $r_{\text{res}} = \SI{1.05(15)}{\milli\meter}$ was estimated from the Siemens star images (\mbox{fig. \ref{fig:resultsSiemensStar} (c-f)}). This resolution is not far from the maximum we can expect if we compare it with the full width at half maximum of the smallest beam shape \mbox{(fig. \ref{fig:resultsTHzTDSbeamProfiling} (d, e))} of $\SI{0.65(10)}{\milli\meter}$ and the wavelength of  maximal intensity of the THz-TDS ($\SI{0.6}{\milli\meter}$ corresponding to \SI{0.5}{\tera\hertz}), which determines the maximal achievable resolution. We presume that one relevant contribution to the limited resolution is the use of a broadband emitter instead of a single line source. Although the radiation contains high frequencies, which would allow for better spatial resolution, the dominant signal from lower frequencies blurs the image and dominates resolution properties. Furthermore, the strong water vapor absorption under ambient conditions greatly reduces the intensity of shorter wavelengths, leaving only longer wavelengths available for imaging. Since the image quality highly depends upon the optical path length, reducing the overall distance between source, sample and camera in conjunction with suppressing the water vapor influence would help. Not only if working over larger distances or under ambient conditions is absolutely required, the ongoing development of more efficient PCAs could provide a larger frequency bandwidth and higher intensities at shorter wavelengths. Blocking the lower frequencies with a high pass filter whilst still maintaining enough intensity for a high contrast THz image would allow resolving smaller structures which are not accessible with the current setup.

\subsection{Potential applications}
	\label{sec:possibleApplications}
	For the THz community, real-time beam profiling of weak and/or broad band sources like shown in \mbox{figure \ref{fig:resultsTHzTDSbeamProfiling}} is still a difficult task. The results shown in this work also suggest that the new generation of very sensitive THz microbolometer cameras can image beams emitted from a PCA, but also other weak THz sources. 
	\newline With the proof-of-concept that complex sample structures can be imaged (\mbox{fig. \ref{fig:resultsLeaves} \& fig. \ref{fig:resultsWood}}), the setup described in this work can also provide material scientists with easier access to THz experiments. 
	\newline We can think of experiments with polymers, aerogels, (embedded) nanomaterials and materials derived from modified biological precursors. The latter class can also act as a bridge to biology. In-field, in-vivo experiments on leaves, grass, crops, young (tree) saplings are all within the realms of possibility. Due to the comparatively high portability, low power consumption and robustness, long term field experiments in remote locations without infrastructure/off-grid seem feasible.
	\newline From the capability to image biological samples, the agriculture and food industry could also profit directly, 
	\mbox{e.\,g.} improving water management by monitoring the water content of leaves and plants. This has been demonstrated before with various THz setups \cite{gente_monitoring_2015}. Not only could this imaging method be applied during the production, but also ensure the product quality during transport and further processing, \mbox{e.\,g.} real-time detection of spoiled products or foreign bodies through packaging (see \mbox{fig. \ref{fig:resultsKey}}). The latter one also immediately implies suitability for security applications like mail screening.  
	\newline  Further industrial use cases could be quality control during production, \mbox{e.$\,$g.} monitoring of the water content in paper, safety inspection or recycling of plastics, etc. Also thinkable would be THz photoelasticity, where one measures the stress states of THz-transparent materials in transmission.
	\newline
	The broad spectral range and polarization control can be used to visualize and evaluate residual stress distributions in packaging materials and electronic housings, but also to display stress distributions in real-time during a mechanical test.

\subsection{Outlook: Possible modifications}
	Since imaging in the \THzTDS zigzag geometry has been shown (\mbox{fig. \ref{fig:resultsSiemensStar}} (b, c)), 
	imaging using a PCA is not restricted to a conventional lens-based THz microscope. Instead of using lenses, setups with OAPMs specially designed for imaging are possible. This could allow accessing methods/samples which cannot be used/measured with commercially available (macro) lenses. A major disadvantage for achieving good image quality with this concept are the many degrees of freedom of an OAPM. This makes precise alignment very challenging, but exact calibration is an essential prerequisite for obtaining a high quality image. 
	\newline During our initial experiments with the new source-detector combination, we focused on transmission. 
	However, it is straightforward to change the setup from transmission to reflection imaging by rearranging the illumination geometry.
	\newline  To mitigate the loss of time domain/spectral information caused by using a microbolometer camera instead of a conventional RX, we propose multi-spectral imaging via inserting THz band pass filters into the beam path. Mounting the filters onto a rotating wheel could allow the generation of false-color images in real-time. Its feasibility depends upon the quality of THz filters and the availability of PCA sources with higher power.

%% file: sections/pca-imaging_conclusion_v1.tex
We presented a THz real-time~imaging method based upon a~fiber coupled~PCA and an uncooled~microbolometer camera. As a proof-of-concept, we recorded a \THzTDS beam shape in real time and examined the performance for typical security, quality control and material science tasks in transmission geometry. Challenges encountered during the experiment were the weak sample irradiance, a resolution lower than the physical wavelength limit and for some samples reduced image quality. Furthermore, we discussed possible improvements and (practical) applications of the setup, including experiments in reflection geometry and multi-spectral imaging with THz filters.